\documentclass[conference]{IEEEtran}
\IEEEoverridecommandlockouts
\usepackage{cite}
\usepackage{amsmath,amssymb,amsfonts}
\usepackage{algorithmic}
\usepackage{graphicx}
\usepackage{textcomp}
\usepackage{xcolor}
\usepackage{booktabs}
\usepackage{cleveref}
\usepackage{url}
\usepackage{flushend}
\usepackage{subfigure}
\usepackage{xspace}

\def\BibTeX{{\rm B\kern-.05em{\sc i\kern-.025em b}\kern-.08em
    T\kern-.1667em\lower.7ex\hbox{E}\kern-.125emX}}

\newcommand{\toolname}{PipeSim\xspace}
\newcommand{\aiops}{AI ops\xspace}

\begin{document}

\title{{PipeSim}: Trace-driven Simulation of Large-Scale\\{AI} Operations Platforms}

\author{\IEEEauthorblockN{Thomas Rausch}
\IEEEauthorblockA{\textit{TU Wien / IBM Research AI}\\
t.rausch@dsg.tuwien.ac.at}
\and
\IEEEauthorblockN{Waldemar Hummer}
\IEEEauthorblockA{\textit{IBM Research AI}\\
waldermar.hummer@gmail.com}
\and
\IEEEauthorblockN{Vinod Muthusamy}
\IEEEauthorblockA{\textit{IBM Research AI}\\
vmuthus@us.ibm.com}
}

\maketitle

\begin{abstract}

Operationalizing AI has become a major endeavor in both research and industry.
Automated, operationalized pipelines that manage the AI application lifecycle
%---from model training, to runtime performance monitoring, to automated retraining---
will form a significant part of tomorrow's infrastructure workloads.
%AI workflow platforms will be challenged to efficiently operate the vast number of AI
%pipelines as the number of AI users continues to increase.
To optimize operations of production-grade AI workflow platforms we can leverage existing scheduling approaches,
yet it is challenging to fine-tune operational strategies that achieve application-specific
cost-benefit tradeoffs while catering to the
specific domain characteristics of machine learning (ML) models, such as accuracy, robustness, or fairness.
We present a trace-driven simulation-based experimentation and analytics environment that allows researchers
and engineers to devise and evaluate such operational strategies for large-scale AI workflow systems.
Analytics data from a production-grade AI platform developed at IBM are used to build a comprehensive simulation model.
Our simulation model describes the interaction between pipelines and system infrastructure,
and how pipeline tasks affect different ML model metrics.
We implement the model in a standalone, stochastic, discrete event simulator, and
provide a toolkit for running experiments.
Synthetic traces are made available for ad-hoc exploration as well as statistical analysis of
experiments to test and examine pipeline scheduling, cluster
resource allocation, and similar operational mechanisms.

\end{abstract}

\section{Introduction}
\label{sec:intro}

\begin{figure}[b!]
	\centering
	\includegraphics[width=1.0\columnwidth]{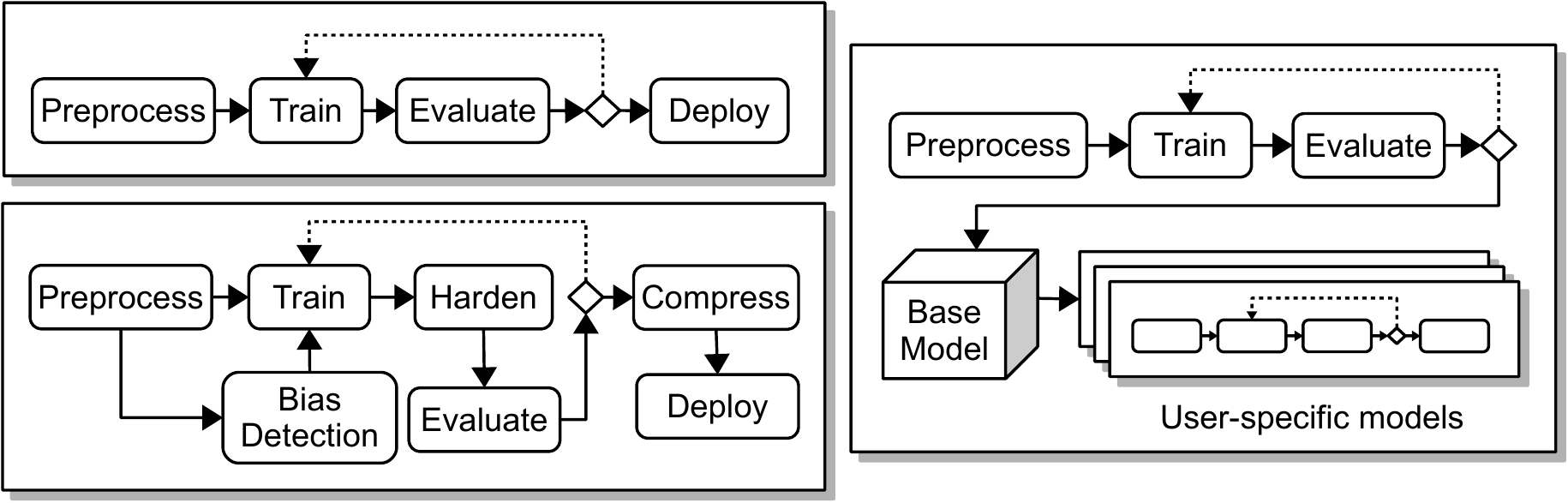}
	\caption{Prototypical AI pipelines: (1) simple process-train-validate-deploy workflow,
  (2) extended pipelines with custom steps, (3) hierarchical pipelines with transfer learning}
	\label{fig:pipelines}
\end{figure}

Developing and operating artificial intelligence (AI) applications involves complex workflows that comprise many tasks performed in an increasingly automated fashion.
This includes data preprocessing flows, data bias checks, machine learning (ML) model training, vulnerability mitigation algorithms, model compression steps, and potentially many others.
Some examples of such workflows are shown in Figure~\ref{fig:pipelines}.
Systems such as ModelOps~\cite{hummer2019modelops} are used to compose and automate such high-level AI workflows into pipelines that track both build and runtime metrics of deployed ML models.
Pipelines are re-executed either automatically due to events such as arrival of new training data, or manually by developers or data scientists~\cite{wang:17, zeager:2017}.
Furthermore, these pipelines can be long running.
It is not uncommon for the training of a deep neural network, for example, to take days.
Consequently, a production system may have hundreds or thousands of pipelines executing at any given time to build and maintain AI models.

In one of the use cases we have explored as part of this research, a company from the health care domain trains ML models to predict certain health conditions of their patients based on real-time monitoring of bodily functions.
The re-training has traditionally been performed manually every four weeks with the newly collected data (which are in the double digit millions).
Manual maintenance of these pipelines turned out to be unsustainable, especially as the company is moving to shorter training intervals and training of large numbers of patient-specific models.
%
%System support for executing these AI pipelines has become a major endeavor in both research and
%industry. Automated, operationalized pipelines that manage the AI model and application lifecycle,
%from model training, to runtime performance monitoring, to automated retraining, will be a
%significant part of tomorrow's infrastructure workloads.

AI platforms need to cope with an ever increasing number of AI pipelines used to manage the AI application lifecycle.
Capacity planning becomes more difficult given the heterogeneous workloads and infrastructure required,
and the variety of automation rules that trigger pipeline executions.
Developing operational strategies, such as optimized scheduling of automated pipelines, therefore plays a major role in the development of AI platforms.
Ultimately, our goal is to leverage the runtime data of pipeline executions to build
predictive models that allow us to automatically derive optimized scheduling decisions
for continuous improvement of AI models.
To that end, this paper contributes a trace-driven simulation framework, for analyzing and experimenting in large-scale AI operations platforms.
This paper is an extended version of the OpML'20 paper~\cite{rausch2020pipesim}.
Our approach builds on modeling the structural and qualitative aspects of AI pipelines, and executing these pipelines in a simulated environment, to fine-tune operational strategies in the real system, by systematically mutating parameters in an iterative, exploratory process.
The simulation model reflects existing high-level AI ops platform architectures such as ModelOps~\cite{hummer2019modelops} and IBM Watson OpenScale,
and considers concepts from AI application workflow as first-class citizens.
We implement the model in a stochastic discrete event simulator using existing techniques \cite{fishman:13,simpy:2008}.
To obtain a realistic representation of different execution parameters (e.g., job arrival patterns, task execution times, training data sizes), we apply clustering and sampling techniques to extract statistical distributions from a database of real-world usage events of a large-scale cloud-based AI platform.
The database contains millions of data points recorded from several thousand of pipeline executions over the course of over a year.
%The core novelty of our approach is that it goes beyond resource-centric job scheduling,
%and considers performance metrics of AI models (e.g., prediction confidence, model
%robustness\cite{weng:18}, model fairness) as first-class citizens of the operational strategies.
%To that end, we introduce the notion of \emph{model staleness}, which expresses a
%cost-benefit tradeoff between the costs of continuously monitoring and improving a
%model versus the risk of the model becoming inaccurate, vulnerable to attacks, or biased.
To the best of our knowledge, this is among the first papers that presents a comprehensive
system model, data acquisition approach, and experimentation environment for large-scale
AI operations platforms that span across the entire lifecycle of AI models.

Our evaluation shows that the framework provides critical and actionable insights into the
performance of AI lifecycle pipelines in terms of system utilization, resource bottlenecks,
and the impact of different platform usage patterns.
% \todo{we can't really back this next claim}
%as well as the system impact of model performance degradation over time.
Providing a link that reconciles from the experimentation environment to the real system,
the approach enables us to evaluate operational strategies and fine-tune pipeline scheduling
algorithms in the platform and do capacity planing.
Furthermore, we show that our approach can simulate and analyze years of data on
a standalone machine in reasonable time, and that our modeling approach has good
simulation accuracy.

\iffalse
The remainder of this paper is structured as follows. In \Cref{sec:related} we first discuss
related work and put it into perspective with our contributions. We then outline the challenges
of optimized operations of AI platforms, and motivate the research problem in \Cref{sec:background}.
In \Cref{sec:design}, we introduce our system model for AI pipelines, sketch the design of the
experimentation environment, and discuss the approach for acquiring realistic simulation settings from
sampled distributions over real-world data. \Cref{sec:impl} covers implementation details
of the proposed framework, and \Cref{sec:eval} covers an extensive evaluation of the presented
approach. \Cref{sec:concl} concludes the paper with an outlook on future work.
\fi
\section{Related Work}
\label{sec:related}

% FfDL \cite{ffdl:17}

This section discusses related work in the areas of AI lifecycles and operations platforms,
system modeling and simulation, as well as job scheduling for model retraining loops.

%\paragraph{ML training optimization:}
KeystoneML~\cite{keystoneml:2017} optimizes pipelines, but the pipelines capture a
different level of abstraction. It focuses on details of what we abstract as a black-box: training.
\iffalse
\textit{``Practitioners build complex, multi-stage pipelines involving feature extraction,
dimensionality reduction, data transformations, and training supervised learning models
to achieve high accuracy.''}
\fi
Workflows become more complex when the end-to-end lifecycle of a model, including deployment
in production have to be considered.
ModelHub \cite{miao:17} is a platform for lifecycle management
of deep learning models, which uses synthetic datasets simulating a model developer
performing various tasks to develop a face recognition model. While their approach is centered
around enumerating models with different network architectures and hyperparameters, we focus on
the pipeline orchestration layer to derive operational strategies to avoid model staleness and
manage model performance over time.
The idea of optimizing a workload schedule towards ``overall user happiness'' in the machine
learning space has also been explored for problems like multi-tenant model
selection~\cite{li:2018, yu:2018}. This work is focused on maximizing aggregate model accuracy, while
our work can accommodate an arbitrary set of model attributes including training time, accuracy, and
vulnerability scores.

%\paragraph{Model retraining:}
As ML models are based on data, and ``data is expected
to evolve over time''~\cite{gama:14}, it is normal for the predictive power of models to degrade and
require retraining \cite{zeager:2017}. \cite{gama:14} present a taxonomy and survey of approaches to detect and repair
concept drift. These patterns once encoded as AI pipelines can be simulated and optimized with the
platform presented in this paper.
Similarly, recent work on trust in AI emphasizes the need for lineage and traceability of models'
quality attributes across the lifecycle \cite{hind:18}. Asserting model quality requires
system support on the infrastructure and job scheduling level, and our work provides the
foundation to evolve and optimize these operational strategies.

%\paragraph{ML platform:}
TensorFlow Extended (TFX) \cite{tfx:2017} is a large-scale machine learning platform that tackles
the lifecycle of models from data prep to training and deployment.
TFX uses the notion of model \textit{freshness} in the context of retraining. TFX doesn't address the
modeling of lifecycle attributes or simulation to understand and optimize pipelines.

%\paragraph{Simulation:}
There is considerable work on modeling and simulating workflows in different domains. This includes
seminal work on process mining and
simulation of business processes~\cite{liu:2012}, and simulating the operational characteristics of
cloud data centers by platforms like CloudSim~\cite{cloudsim:2011},
iCanCloud~\cite{icancloud:2012},
or recently more platform-specific simulators like All-Spark~\cite{Lin:2018}.
This paper continues in the spirit of this literature but our focus
is on modeling AI pipelines and concepts they operate on.

\iffalse
\textbf{not exactly related work, but i like the sentiment. useful for intro:}
\cite{zeager:2017}
\textit{While certain models may be more successful at detecting fraud than others,
	all are equally susceptible to attack if not retrained.}
\textit{  There exists a cost of retraining the classifier that was not addressed when
	retraining after every round of the game. The cost of retraining could be
	compared to the benefit of changing the classifier, revealing the ideal
	frequency of retraining. This could also be done systematically by
	retraining the classifier only when the AUC falls below a certain value.}
\fi

\section{Optimized AI Platform Operations}
\label{sec:background}

%Developing and operating AI applications involves complex workflows that comprise many
%tasks often performed manually in an ad-hoc fashion. One example is the retraining of
%machine learning models. During the lifetime of an application that makes use of machine
%learning models, developers, data scientists or other experts may decide that it is necessary
%to retrain some or all of the employed models~\cite{miller:2014, zeager:2017}.
%Retraining is then performed on updated data sets, and the new models are manually verified,
%tested, quality controlled, and deployed. This manual workflow is cumbersome and error prone.

%In one of the use cases we have explored as part of this research, a company from the
%health care domain trains ML models to predict certain health conditions of their patients
%based on real-time monitoring of bodily functions. The re-training has traditionally been
%kicked off every four weeks with the new data points that have been collected (which are
%in the double digit millions). Manual maintenance of these pipelines turned out to be
%unsustainable, especially as the company is moving to shorter training intervals and
%training of large numbers of patient-specific models.

We have come across an increasing number of teams that automate their model training and deployment
pipelines. In the following we discuss the relationship between model metrics and AI pipelines,
and the challenges of automating and optimizing the scheduling and execution of~pipelines.

% overall storyline:
% \begin{itemize}
% \item 1. Ad-hoc workflows to build and deploy models don't scale
% \item 2. operationalized pipelines are the answer
% \item 3. how is this done? (pipeline abstractions, workflow execution)
% \item 4. deployed models then get stale
% \item 5. triggered re-execution of pipelines (which can facilitate continuous retraining) will become common
% \item 6. platforms will be challenged to run and manage these workflows in an optimized way
% \item 7. what will be necessary? intelligent scheduling, load balancing, ... in general: operational strategies
% \end{itemize}

\subsection{ML Model Metrics and AI Pipelines}
\label{sec:background:pipelines}

ML model metrics are essential to the AI application lifecycle, as they drive development and deployment decisions.
Several types of metrics have been defined for ML models~\cite{hernandez:2012}.
We distinguish between
\emph{static} (or build-time) metrics of models, which are inherent properties of the model, and
\emph{dynamic} (or run-time) metrics which are attributes that may change over time and depend on the inferencing data.
Static metrics include test accuracy or AUC, ML model size (e.g., number of weights, or bytes), or model robustness (e.g., CLEVER score \cite{weng:18}).
Dynamic metrics include inferencing time, scoring confidence, or concept drift \cite{gama:14}.

A key metric, central to our approach, is model \textit{staleness},
which refers to the decrease in predictive performance of an ML model over time.
A major goal of AI operationalization, and in particular the inclusion of runtime metrics in the decision mechanism for pipeline execution,
is to mitigate negative performance effects in an AI application caused by model staleness.
Based on model staleness, we can synthesize additional metrics that can help us make pipeline scheduling decisions.
For example, we can see staleness as being inversely proportional to the current \textit{potential} of a retraining pipeline to \textit{improve} the model.
This potential could be given by,
a) the current model performance $p(M) \in [0,1]$ of a model $M$,
a composite value that aggregates static and dynamic metrics into a single value,
and b) the newly labeled data available since the last retraining.

%\begin{mydef}
%Given a model $m$, the \emph{staleness} $s(m,t)$ is the actual, or potential reduction in
%performance at time point $t$, expressed as a function of time and number of processed
%(scored) instances.
%\end{mydef}

It is important to notice that this potential improvement captures both, known or measurable quality degradation
(e.g., concept drift, a statistically significant deviation of inferencing data from training data),
as well as unknown or unmeasurable performance risks (e.g., an attacker may attempt to steal a
model by probing it with test requests \cite{tramer:16}).
That means, a deployed model is usually associated with a certain risk of degrading or becoming stale over time \cite{bifet:15}, which is illustrated in Figure \ref{fig:performance-drift}.

\iffalse
\begin{figure}[!htpb]%
	\centering
	\subfigure{%
		\includegraphics[width=0.5\columnwidth]{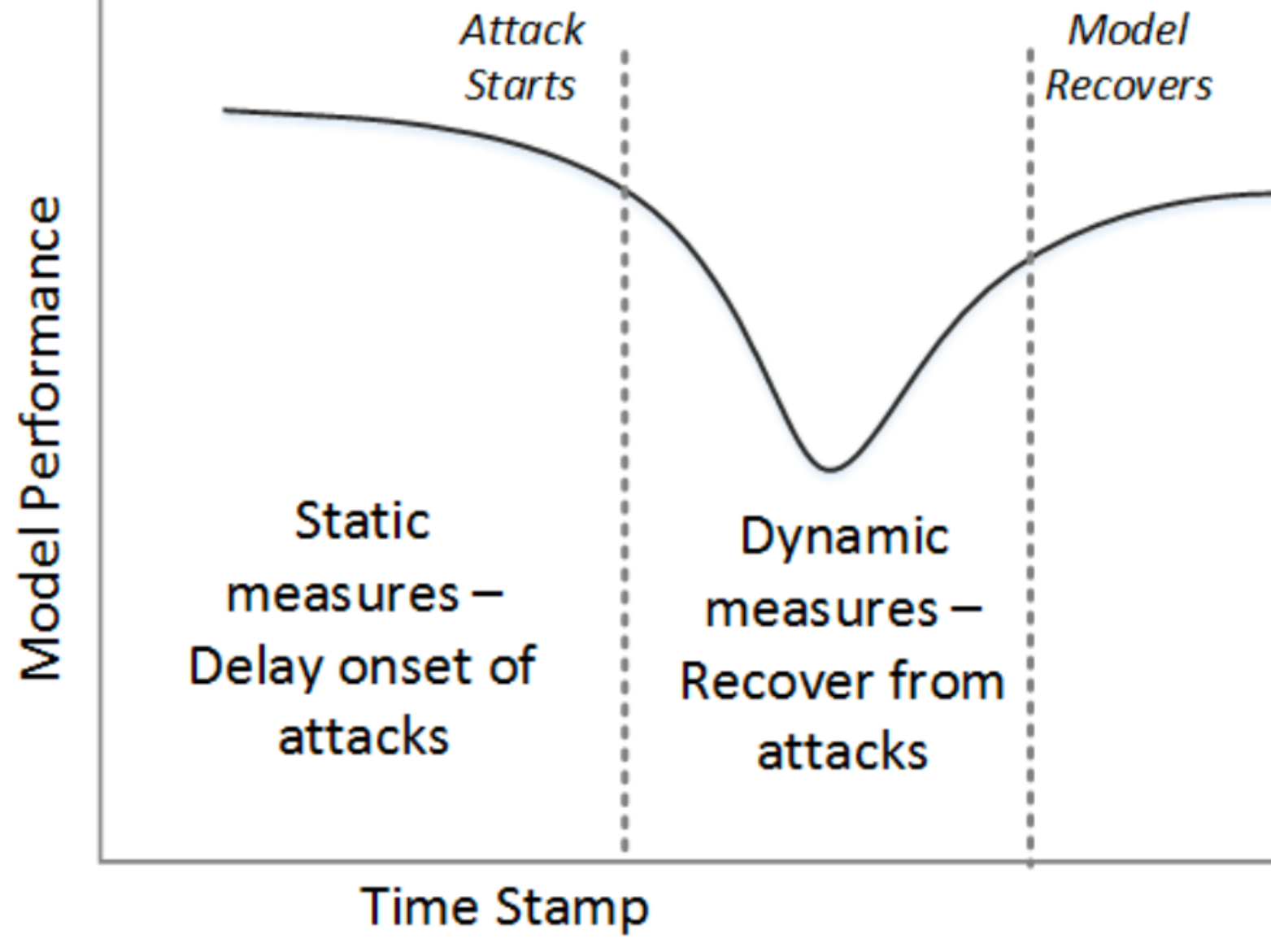}
	}%
	\subfigure{%
		\includegraphics[width=0.5\columnwidth]{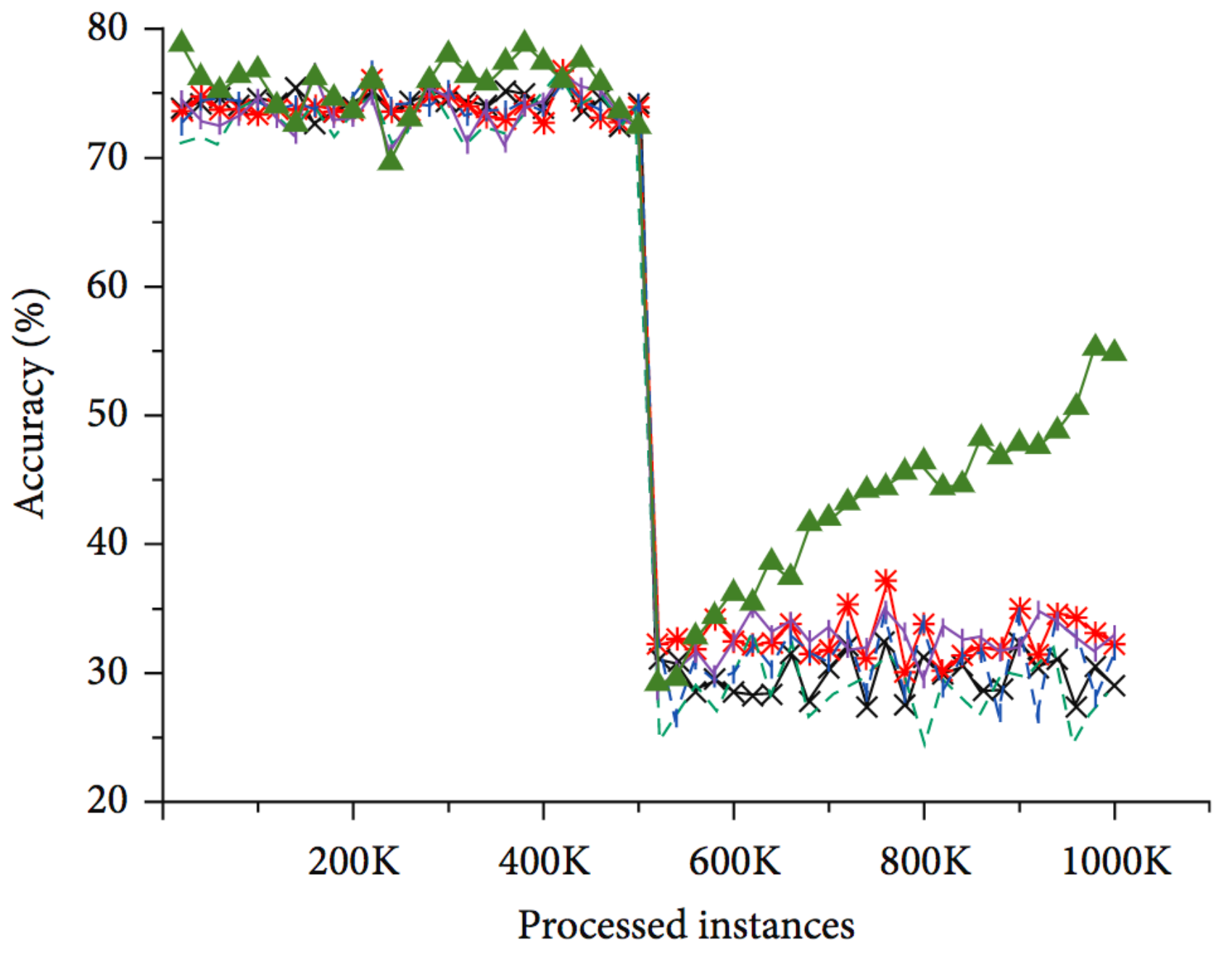}
	}
	\caption{Illustration of model performance over time, indicating an adversarial attack
		\cite{sethi:18} (left), and accuracy drop due to a sudden concept drift \cite{sun:16} (right)}
	%\label{fig:performance-drift}
\end{figure}
\else
\begin{figure}[!htpb]%
	\centering
	\includegraphics[width=0.9\columnwidth]{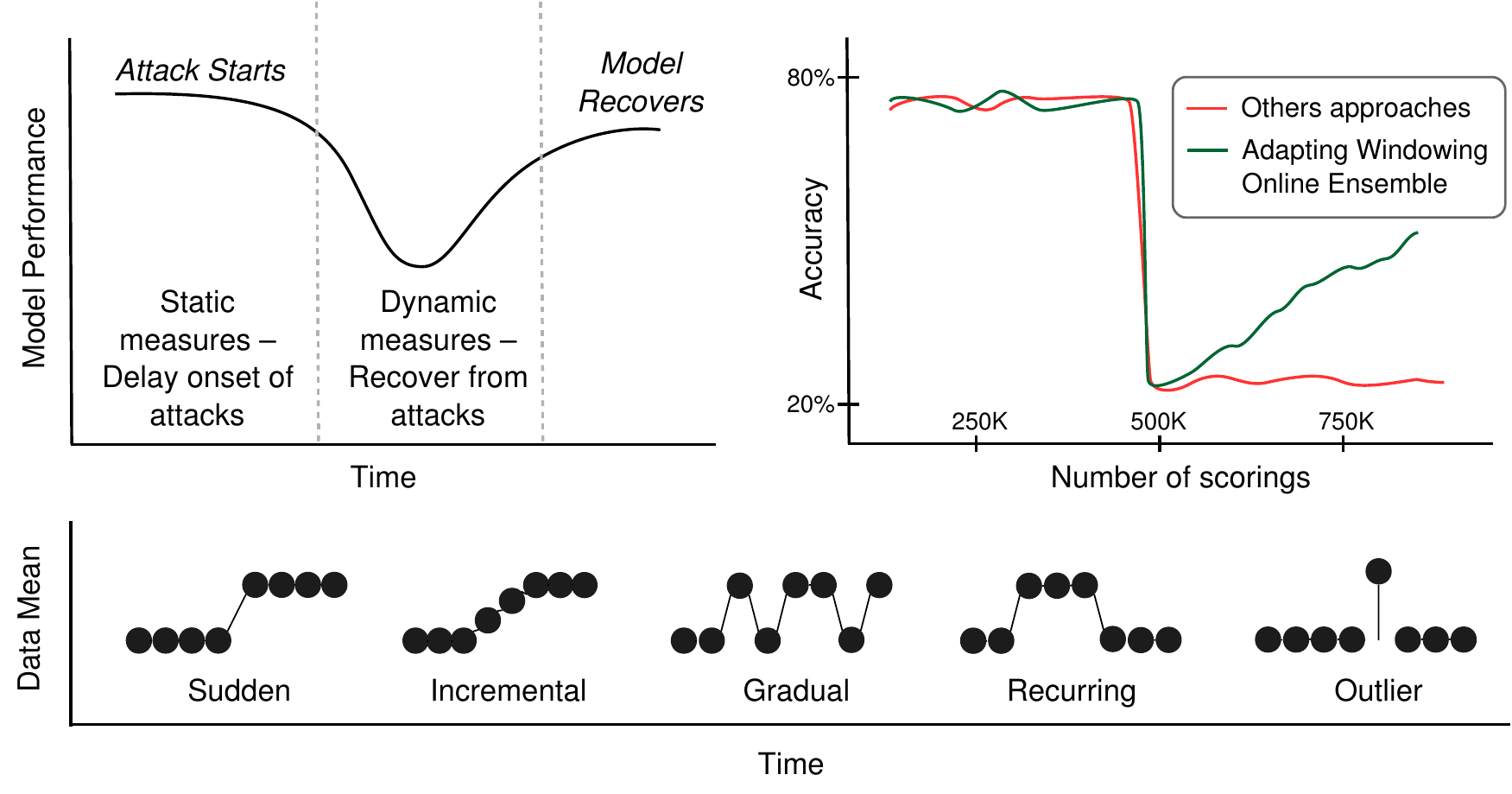}
	\caption{
		Illustration of model performance over time,
		indicating an adversarial attack \cite{sethi:18} (left),
		accuracy drop due to a sudden concept drift \cite{sun:16} (right),
		and abstract concept drift patterns \cite{gama:14} (bottom).}
	\label{fig:performance-drift}
\end{figure}
\fi

% This mechanism allows, for example, the implementation of continuous training loops.

\begin{figure}[!htpb]
	\centering
	\includegraphics[width=0.8\columnwidth]{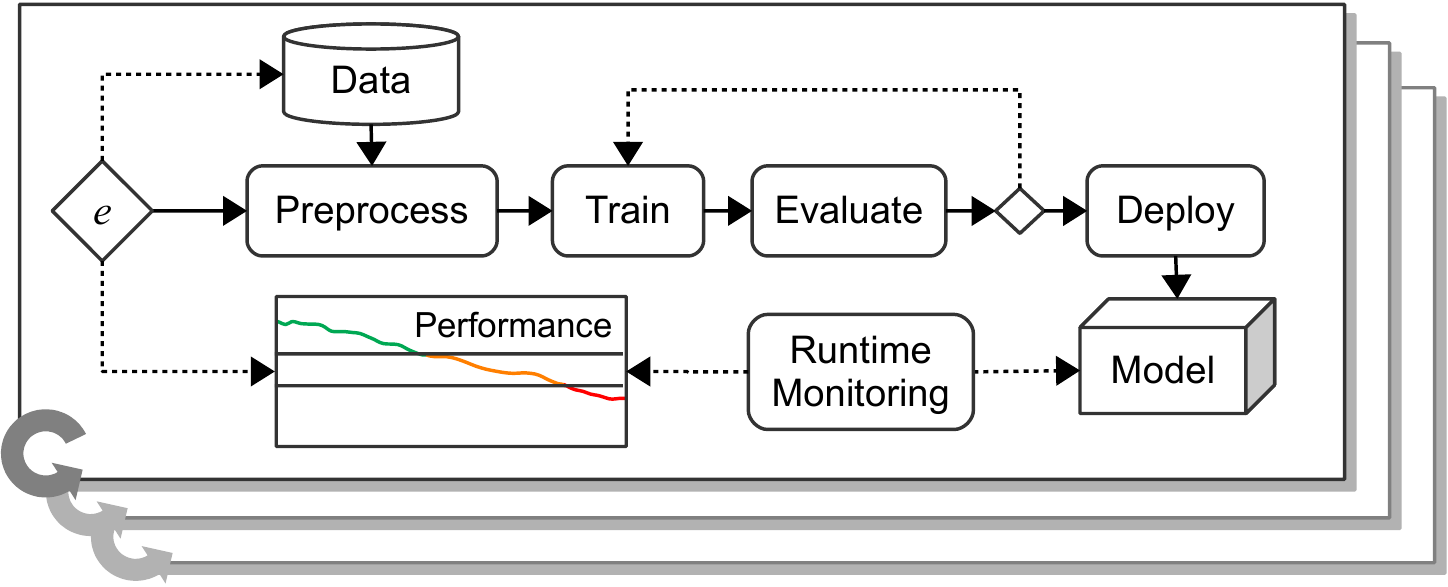}
	\caption{Automated AI pipelines with execution triggers, quality gates, and runtime monitoring}
	\label{fig:automated-pipeline}
\end{figure}

In order to manage the risks and prevent models from becoming stale, it is crucial for AI operations to
employ pipeline automation with triggers that monitor runtime indicators of deployed
models.
% A rule that monitors the number of new labeled data that have arrived since the last training
% and the performance of the model, will trigger a re-training and re-deployment of that model.
An execution trigger $e$ is a set of rules that reason about
the pipeline inputs (such as the dataset used for training, to detect changes or
drift), previous executions of the pipelines (when was the model last built and
deployed), and performance of the deployed model.
If any of the rules underlying $e$ is met, an execution of the pipeline is triggered automatically.
\Cref{fig:automated-pipeline} illustrates the feedback loop.

% \todo{i think the next few sentences may fit better into section 3.1}
The basic unifying characteristic of \aiops pipelines is that they generate or augment AI models.
Each task plays a role in either training a model, or enhancing model metrics such
as performance or robustness.
This concept allows us to reason about a model and its properties, and the pipeline it
is generated by, almost interchangeably.
We can think of the model as a latent component of a pipeline, whose properties are assigned or
changed once its specific tasks are executed.

\subsection{Optimized Operation of Automated AI Pipelines}

\begin{figure}[!htpb]
	\centering
	\includegraphics[width=1.03\columnwidth]{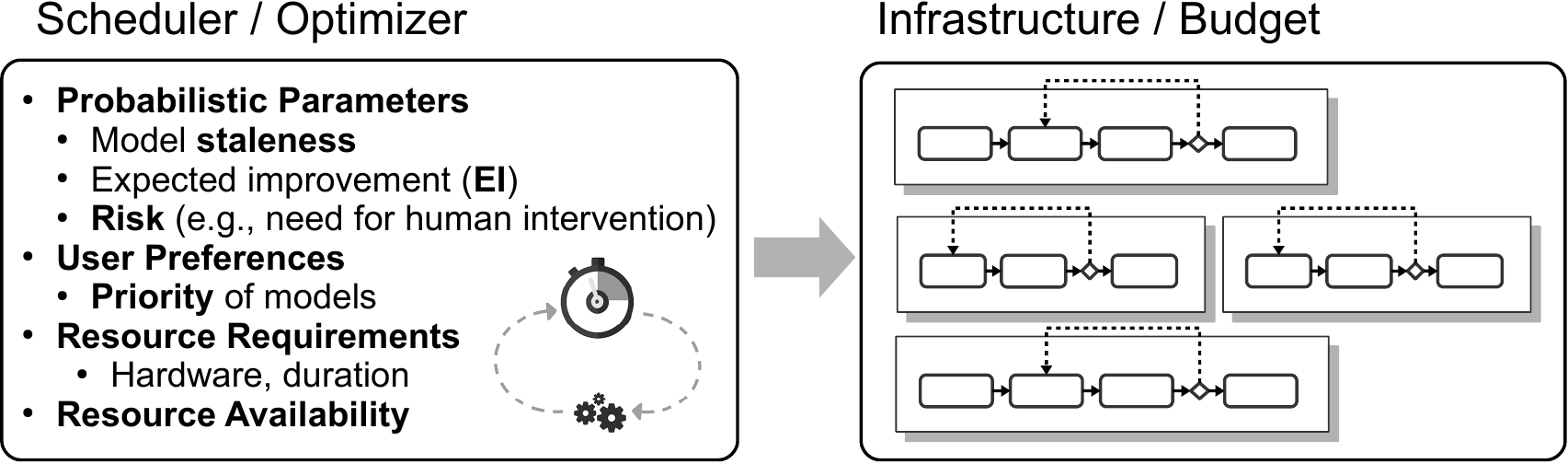}
	\caption{An AI pipeline scheduler optimizes overall user satisfaction and resource balancing}
	\label{fig:scheduler}
\end{figure}

A key challenge is to determine the trade-off of costs associated
with pipeline execution, and the benefit of (expected) performance improvement~\cite{tfx:2017}.
Optimized scheduling and resource sharing for ML pipelines under constraints is an active research
problem \cite{li:2018}, \cite{tfx:2017}.
Previous research has found that ML platform users often do not use infrastructure resources in a useful way\cite{li:2018}.
For example, users would reserve several GPUs for weeks to improve models that already had an accuracy of 0.99.
From a AI platform provider's perspective,
the operational challenge is to reconcile user SLAs (encoded as pipeline execution triggers or rules),
infrastructure utilization,
and fairness.
Given the notion of potential improvement we could conceive a platform that has as goal to
optimize the potential improvement of all automated AI pipelines,
while maintaining SLAs and balancing infrastructure utilization.
Figure \ref{fig:scheduler} conceptualizes a scheduler that deploys pipelines
on to limited infrastructure, based on probabilistic parameters
(e.g., model staleness), user preferences (e.g., model prioritization), and
resource availability.
However, developing and testing such schedulers is difficult given the lack of frameworks and simulation tools, which is what we address in this paper.

% \iffalse
% \fi

% Users can define quality level attributes like model staleness .

% A scheduler, illustrated in \Cref{fig:scheduler} determines the execution of pipelines to
% meet user SLAs, but also to optimize model performance metrics over time. \todo{add more details}

\subsection{Challenges in Operational Research for AI Ops}

Recent work on scheduling and multi-tenant resource sharing allows users to utilize training
infrastructure in an efficient way~\cite{li:2018}. Yet, devising optimal operational strategies is still
challenging due to a number of factors:

\textbf{Cutting Edge}: As large-scale AI pipelines are an emerging field, industry and research have not yet
 	converged to a common configuration format for pipelines, hence pipeline definitions are often
	custom-built and not readily available for analysis or optimization.

\textbf{Limited Testability}: The execution of AI pipelines - including tasks like model training on
	GPU clusters - is often long-running, resource intensive, and hence costly. Operational
	research relies on experimentation with quick feedback cycles, yet executing test workloads to
	evaluate large-scale strategies in the real system may be prohibitively expensive.

\textbf{Lack of Data}: As the field is still evolving, there is currently a lack of datasets
  about AI pipeline executions, capturing the variety of specialized ML tasks (e.g., data preprocessing, model
	training, bias and robustness checks, model compression, etc), including the specific effects of each task on
	the code, data, and model assets, across the entire lifecycle.

\section{Experimentation and Analytics for AI Operations Platforms}
\label{sec:design}

\begin{figure}[!htpb]
	\centering
	\includegraphics[width=1\columnwidth]{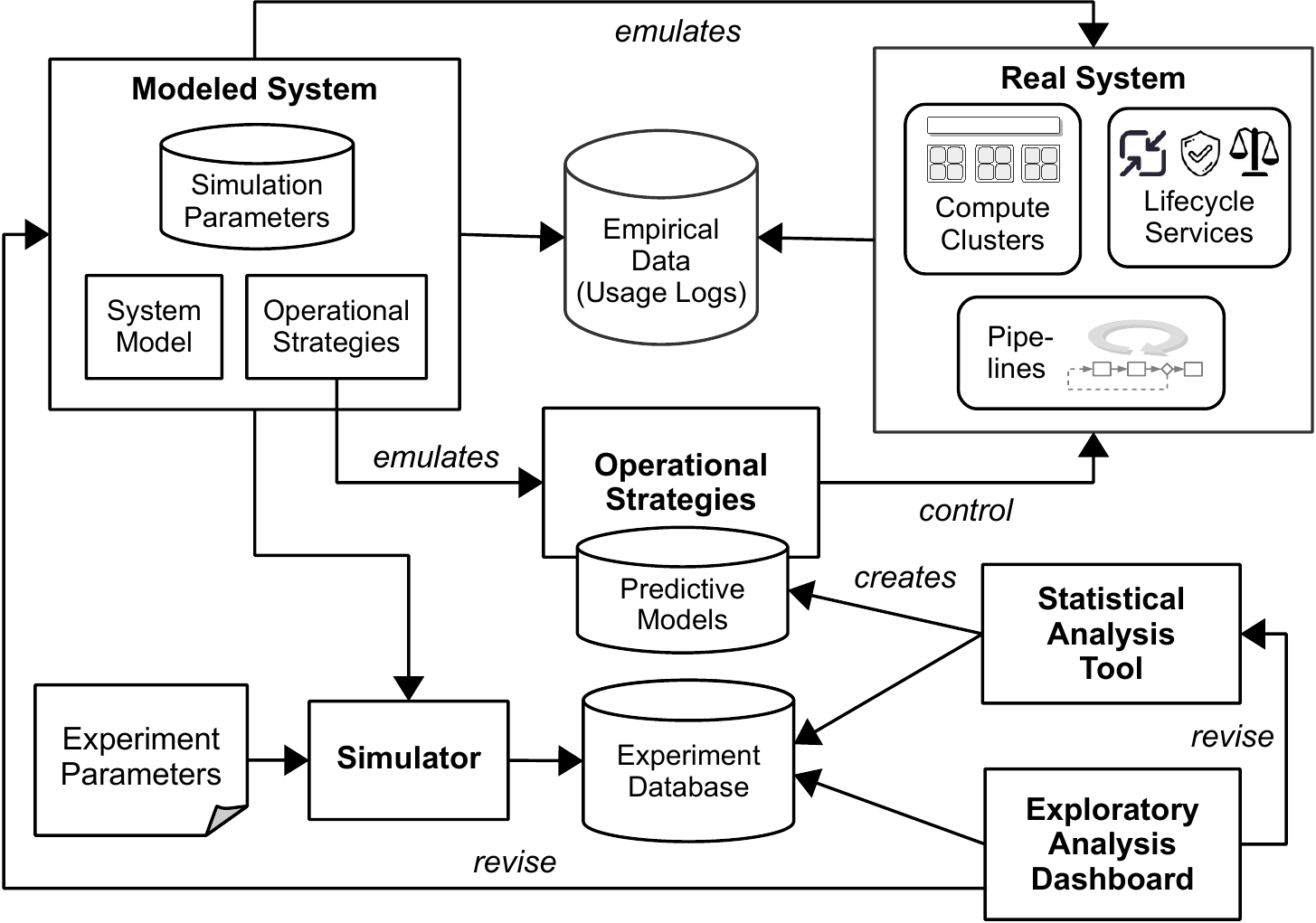}
	\caption{System Architecture of \toolname}
	\label{fig:architecture}
\end{figure}

In this section we introduce \toolname, an experimentation and analytics framework for AI operations
platforms. The system architecture is illustrated in Figure \ref{fig:architecture}. The \emph{real system},
which consists of compute clusters, pipeline executions, and various lifecycle services (e.g., model
compression, model robustness, and data bias checkers), is represented in a \emph{modeled system}, which
feeds into a \emph{simulator} component. The system model is parameterized with simulation parameters whose
distributions are sampled from empirical data, extracted from real-world usage logs of the system.

The \emph{operational strategies} in the modeled system emulate the strategies that control the real
system. For example, a pipeline scheduler would make API calls to the pipeline executor in the real system.
In the modeled system, the scheduler operates on the system model, i.e., creates a pipeline entity
with current timestamp and feeds it into the simulator.

The main entry point for users is to define an \emph{experiment} and its parameters;
once it gets executed by the simulator, the user can drill into
the results using the \emph{exploratory analysis dashboard}. The \emph{statistical analysis tool}
analyzes the results of the experiment database and is used to create predictive models, which
themselves feed into the operational strategies, to close the feedback loop into the real system.

In the following subsections we discuss the overall approach and the individual components of
the system in more detail.

\subsection{AI Ops Platforms: Conceptual System Model}
\label{sec:design:model}

As discussed in \Cref{sec:background:pipelines}, automated \aiops pipelines integrate the
entire lifecycle of an AI model, from training, to deployment, to runtime monitoring.
In our definition of pipelines, we distinguish between the build-time and run-time view of
the platform -- analogous to the two steps in the conceptual
ML workflow described by~\cite{tfx:2017}: the training phase and the inference phase.

\subsubsection{Build-Time View}

\begin{figure}[!htpb]
	\centering
	\includegraphics[width=0.93\columnwidth]{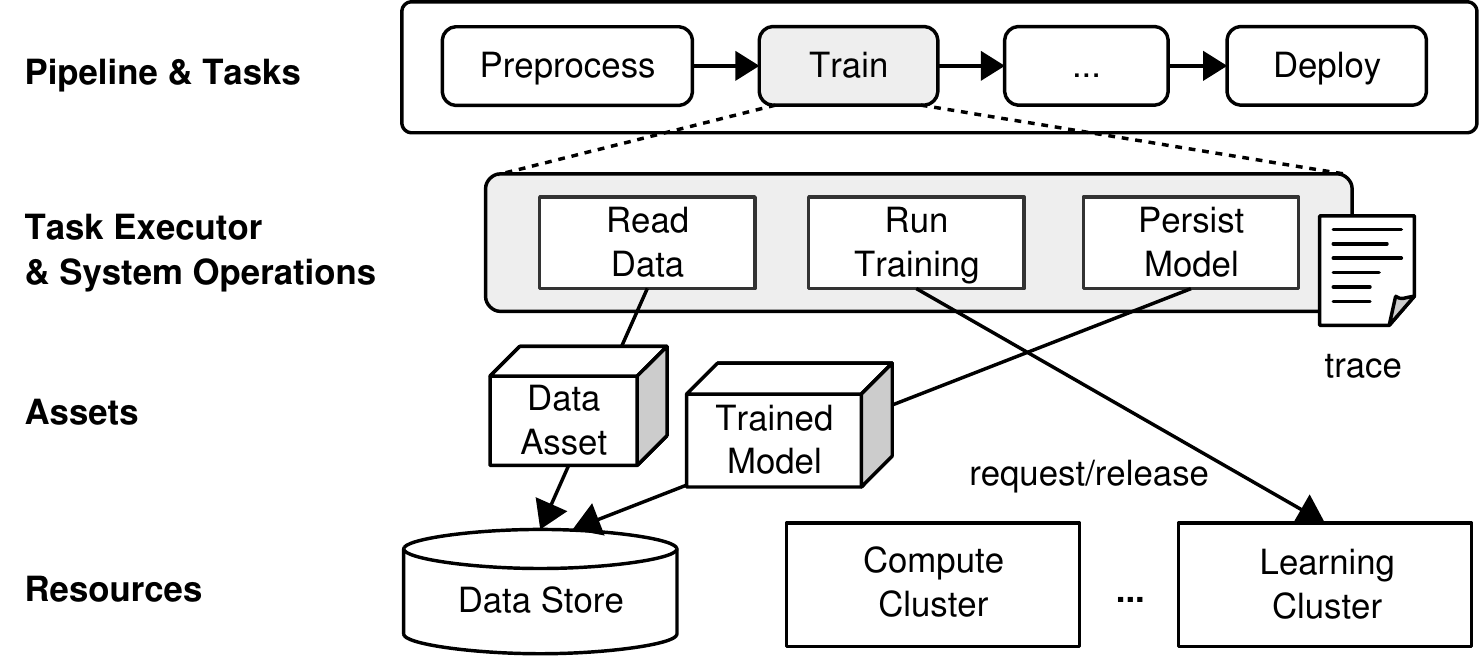}
	\caption{Conceptual system model: Build-time view}
	\label{fig:build-time-model}
\end{figure}

At build time, an AI pipeline generates or augments a model (classifier) by operating
on data assets and using underlying infrastructure resources (e.g., GPUs).
The build-time view of our system model is illustrated in \Cref{fig:build-time-model}.

\paragraph{Pipeline and Tasks}

AI pipelines are compositions of tasks that create or operate on machine learning
models~\cite{keystoneml:2017, tfx:2017}.
Formally, a pipeline $G_{p}$ is a directed graph (digraph) $ G_{p} = (V_{p}, E_{p}) $,
where vertices $V_{p}$ are tasks (i.e., workflow operations),
and directed edges $E_{p}$ are task transitions labeled with the input that triggers a transition.
To better reason about the control flow it is useful to also explicitly model decisions and joins
(task that are only transitioned to if all previous task have been executed).
However, as this is not the focus of the paper, we omit these definitions.
We model different types of tasks denoted $\tau$, where
$ \tau \in \{
	\mathrm{preprocess},
	\mathrm{train},
	\mathrm{evaluate},
	\mathrm{compress},
	\mathrm{harden},
    ...
\} $.
We shorten the notation to the first letter of the type.
Instances of tasks of a specific type in a pipeline are denoted by
$v^\tau \in V^\tau_p \subset V_p$. Each task instance holds type specific variables,
for example we associate the original and transformed data asset $D$ and $D'$ with
a data preprocessing task~$v^p$.

\paragraph{Resources}

A resource $R$ represents an infrastructure component required for executing pipeline tasks.
We model a generic system comprising (i) a generic data storage, (ii) a training, and (iii) a compute infrastructure,
but we allow for custom configuration fo resource types.
Data stores are abstracted in terms of read and write operations, and can therefore be anything from a cloud-based object store (such as S3), to a relational or NoSQL database.
% All a Data Store has to provide are generic read and write operations that operate on Assets.
% Compute resources model infrastructure where specific task operations are executed on, such as a compute or learning cluster.
Cloud-based AI platforms, such as Watson OpenScale,
perform training of models on a dedicated infrastructure with specialized hardware (GPUs),
whereas generic compute tasks, such as data preprocessing, may be performed on general purpose compute hardware (e.g., running Spark or Hadoop).
%We assume a learning infrastructure that is framework agnostic, such as FfDL~\cite{ffdl:17}.
Each compute resource is assumed to have a specified job capacity and a work queue,
but we do not make assumptions about individual scheduling mechanisms.

\paragraph{Assets}

An asset $A$ denotes any data artifact that is transferred, processed, and generated by data stores
and compute resources.
% play a key role in the \aiops workflow. Modeling of assets, in particular their size, is
% therefore essential for implementing realistic simulation of pipeline task executions.
Modeling asset characteristics is critical,
as the execution time of pipeline tasks often significantly depends on the size and dimensions of processed data sets,
and is necessary to simulate traffic between tasks and data stores.
We distinguish between Data Assets $D$ and Trained ML Models $M$.
% Furthermore, knowing the disk space requirements of a data asset is essential to model traffic
% and bandwidth consumption between tasks and a data store.
%In general, we model an asset to be anything that has a size in bytes.
%For data assets we model two additional key characteristics:
%number of dimensions (columns), and rows.
%A Trained Model Asset holds additionally meta-data about the model and estimator type.
This allows us to store ML model specific characteristics, like the number of neurons or layers of neural networks.

\paragraph{Task Executors}

Task executors encapsulate the actual system operations performed by a task.
For example, a task executor of a training task on a distributed compute cluster
%such as FfDL~\cite{ffdl:17},
is an iterative process of fetching training data from a remote storage (like S3) and
running an optimization algorithm (e.g., gradient descent) on the data,
and finally persisting the model to the storage.
We define the following system operations
$ \Omega = \{
	\mathrm{read}(A),
	\mathrm{write}(A),
	\mathrm{req}(R),
	\mathrm{rel}(R),
	\mathrm{exec}(v^\tau_p, R)
\} $
Where the first to are read and write operations on the data store for an Asset $A$,
$\mathrm{req}(R)$ and $\mathrm{rel}(R)$ are request and release operations for compute resources,
and $\mathrm{exec}(v^\tau, R)$ is the task type $\tau$ specific execution of
$v^\tau_p$ on $R$ which describes how that task interacts with the system resources.
A task executor is therefore a sequence of operations $ (\omega_0,...,\omega_n), \omega_{i} \in \Omega $.
Typically the first and last operations in a sequence are read and write operations, respectively.
% This is description is general enough to encompass a variety of systems, but specific
% enough to analyze the impact of executing a training task on the system in a meaningful way.
%Task executors generate execution traces that we can later analyze.

\subsubsection{Run-Time View}

% \todo{a little thin}.
The outcome of a successful pipeline execution is usually a deployed model
that is being served by the platform and used by applications for scoring.
At runtime, the deployed model has associated performance indicators that change
over time. Some indicators can be measured directly, by inspecting the scoring
inputs and outputs (e.g., confidence), whereas other metrics (e.g., bias, or drift)
require continuous evaluation of the runtime scoring data against the statistical
properties of the historical scorings and training data.

\begin{figure}[!htpb]
	\centering
	\includegraphics[width=\columnwidth]{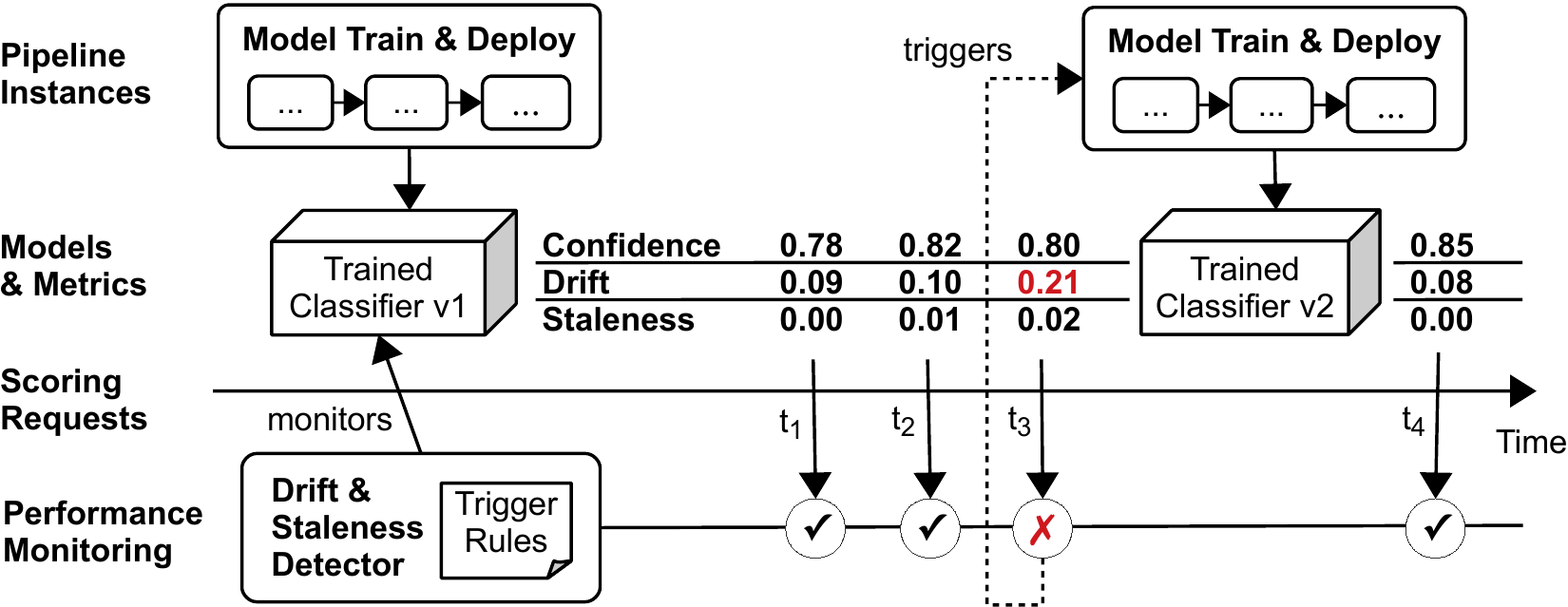}
	\caption{Run-time view: Drift detection and continuous retraining}
	\label{fig:drift-detector}
\end{figure}

% Because we associate with the deployed model the pipeline that it was generated by,
% we can also reason about, e.g., the training data that generated the model,
% and include metrics like concept drift.
% From a system's perspective, it is also interesting to model the size of the data that
% is sent for inferencing to determine the traffic generated by a deployment.
The ability to react to scoring events, introspect the historical payloads, and continuously
compute performance metrics is a critical piece of functionality that itself requires a
significant amount of computational resources. In fact, model performance monitoring tools
like drift detectors are themselves ML models, for example based on model
explanation classifiers like IME \cite{demvsar:18}, that need to
be trained, deployed, and continuously maintained.

Figure \ref{fig:drift-detector} illustrates
the interaction between model assets and performance monitoring in \toolname.
A detector component monitors a trained classifier, and computes drift and staleness metrics
over time. At time point $t_3$, a trigger rule detects that the drift metric exceeds a
threshold, and triggers a retraining pipeline which creates version $v2$ of the
classifier. Note that the pipeline may employ active learning with humans in the loop
to perform data selection and labeling.

% In \toolname we model these as additional
% artifacts as pipeline tasks and assets, which contribute to the resource consumption and
% operational strategies. \todo{briefly describe Figure}

\subsection{Pipeline and Data Synthesizer}
\label{sec:design:datasynth}

To run an experiment we need to artificially generate data that follow some underlying process we can control.
We describe our approach to generate synthetic pipelines and assets that are used as input for a simulation.
%It is desirable for synthetic data to accurately reflect properties and frequencies of the system under simulation.
%However, we also want to be able to perturb data parameters to answer ``what-if'' questions about the system under different conditions.
%For example, we could be interested in the resource utilization of clusters given different pipeline compositions,
%or observe the effect of data asset sizes increasing.
%\todo{flow}

\subsubsection{Synthetic Pipelines}

Because the goal is to simulate the execution of a large number of pipelines,
a key element of the experimentation environment is a Pipeline Synthesizer that stochastically
generates plausible pipelines.
That is, although there should be some randomness involved, the sequence of steps in a synthetic
pipeline should be sensible (e.g., a model validation task cannot precede a training task).
This process is challenging because the structure of complex \aiops pipelines, i.e., beyond simple
train--deploy workflows, is still poorly understood.
However, we can make some basic assumptions about pipeline structures
based on the prototypical pipelines we have identified by analyzing both commercial
and research use cases as presented in \Cref{fig:pipelines}.

We also recognize that some steps within these pipeline structures may be optional.
While a pipeline that generates a model unconditionally requires a training step,
not all require a data preprocessing step if they make use of already structured and
curated datasets. When generating pipelines, some tasks therefore have a certain
(possibly conditional) probability associated with them,
that may depend on the state of the pipeline currently being generated.

Task characteristics, such as the framework and algorithm used for training,
or the number of operations in a preprocessing step, are also synthesized.
Such characteristics can be based on frequencies observed in production systems,
or configured as experiment parameters.
For example, while examining the AI pipeline executions of our platform
we found that
63\% of training jobs use SparkML,
32\% use TensorFlow,
3\% PyTorch,
1\% Caffe, and 1\% use a variety of other frameworks.
Given that different frameworks map to different infrastructures
(e.g., a Spark cluster) and often correlate with significantly different
execution times (see later in Figure \ref{fig:compute-training}), we want to
easily adapt these percentages to observe the effect on the system.

\subsubsection{Synthetic Assets}

A key ingredient of AI pipelines are assets, i.e., training data assets, and
trained models. We describe properties of assets as multivariate random variables,
as this allows us to use common sampling methods for synthesizing assets.

For example, we model a data asset $D$ as an observation of a multivariate
random variable $\mathbf{D} = ( D_d, D_r, D_b )$ where
$D_d$ constitutes the number of dimensions of the dataset (e.g., number of columns in tabular data),
$D_r$ the number of rows or instances, and
$D_b$ the disk space in bytes required to store the asset (uncompressed).
In \Cref{sec:impl:stats} we describe how we can obtain synthetic data assets
by sampling from a multivariate Gaussian mixture fitted on empirical data.

Describing a trained model $M$ in this way is not as straight forward, as many of its properties
are the result of stochastic processes in the pipeline execution.
%\todo{not sure how to describe this here. this is a cross cutting concern}
In general, we say that a model has a set of \textit{static} and \textit{dynamic} properties.
Static properties include those that are assigned by the pipeline at build-time,
such as
the prediction type $M_t$ (e.g., binary or multiclass classification, or regression)
or the model type $M_e$ (e.g., Linear Regression, Random Forest, or Neural Network).
Dynamic properties include metrics we have described in Section~\ref{sec:background:pipelines},
such as model performance $M_p$, and CLEVER score $M_{\mathrm{CLEVER}}$\cite{weng:18}.

\subsection{Process Simulator}

\subsubsection{Task Execution}

Besides the side effects a task execution has on the properties of an Asset,
we are mostly interested in simulating the execution duration of a task.
Because a task is a sequence of system operations $(\omega_n)$,
we can define the execution duration of a task $t(v^\tau)$
as the sum over the execution duration of the task's system operations, i.e.,
$t(v^\tau) = \sum t(\omega_i) $.
Similarly, because we define a pipeline as a sequence of steps,
we can define the execution duration of an entire
pipeline as the sum over task execution durations (the current system model assumes that
tasks are not running in parallel).
This allows granular modeling of system behavior.
For example,
$t(\mathrm{req}(R))$ equates to the time a task as to wait for a resource $R$ to become available,
$t(\mathrm{read}(A))$ and $t(\mathrm{write}(A))$ are functions of $A_b$ and the up/download bandwidth and latency associated with the data store,
and so on.
These functions can be expressed as analytical functions of system properties.
However, for functions that are subject to complex processes,
such as
data preprocessing $t(\mathrm{exec}_p(v^p, R))$,
or model training task $t(\mathrm{exec}_t(v^t, R))$,
we rely on empirical data and statistical modeling of these function.

\subsubsection{Pipeline Arrivals}

A fundamental parameter of a platform simulation is the rate at which pipeline executions are triggered.
We say the average arrival rate is $\lambda$.
A typical way for simulating arrivals is to model the time between arrivals (interarrivals) as a random variable \cite{simpy:2008}
and sampling from the underlying distribution after every event.
It is well known that interrarivals typically follow some exponential or related process.
Researchers have found that, for example, TCP traffic is well described by lognormal, Pareto, or Weibull distributions \cite{feldmann:2000}.
Our data suggests that pipeline-interarrival-times $\delta$ follows a exponentiated Weibull process (see \ref{sec:impl:stats}).
However,
pipelines are executed at a given time either because
they were triggered manually by a user (or other application),
or they were triggered automatically due to a rule.
To us as an operator,
the former is a random process,
whereas the latter we have control over.
It is therefore useful to simulate the processes that underly the user-defined rules
(e.g., the arrivals of new data, the expected model drift, etc.).
Describing these processes is part of the run-time view of the system and is ongoing work.
Furthermore, it is important to preserve, to some degree, the complexity of workload arrival patterns,
and there are many ways to achieve this.
We discuss our approach in more detail in \Cref{sec:impl:stats}.

\section{AI Platform Simulator}
\label{sec:impl}

This section presents our effort towards a full implementation of the experimentation and analytics environment as described in the previous section.
The current implementation comprises a simulator backed by empirical data from \textit{(product name blinded)},
and an analytics frontend connected to a time series database that contains the simulated system data.
% Further features, such as a component that learns usage patterns from the simulation data are work in progress.
In this section we first describe how we acquired the data for statistical modeling of simulation processes.
We then describe the implementation of individual components.

\subsection{Statistical Modeling and Pipeline Simulation}
\label{sec:impl:stats}

As discussed in~\Cref{sec:design:datasynth},
a fundamental requirement for an experimentation environment is to allow meaningful reasoning over the original system,
and therefore requires generated data and underlying process simulation to reflect properties of that system.
While we can make some assumption about, e.g., the distribution of number of pipelines per user
(where the Pareto principle will likely apply),
most data should be based on empirical observations.
The analytics database of \textit{(product name blinded)} with several million rows of user
and system events is our primary source of empirical data.
We run queries on this database an fit different statistical distributions on the extracted data,
which include several thousand pipeline execution traces.

In general, we generate random entities in the simulator by first using scikit-learn or SciPy to fit statistical models on the respective observed data.
The generated models or distribution parameters are exported using Python's serialization to the simulator.
During simulation time, we can then sample individuals from the statistical model.
In principle, fitting all necessary models that we currently use could also be done on the fly
when starting the simulation, but would take in the order of a few minutes.
This allows us to plug in the live, updated data sources for grounding simulation parameters.

\subsubsection{Synthesizing Data Assets}

The data processing component of \textit{(product name blinded)} stores metadata about
data assets, i.e., the number of dimensions and instances,
as well as the amount of bytes it transfers between object storage and execution platform.
To generate synthetic data assets we sample from the distribution of rows, columns and bytes processed.
Note that, while the usage data we sample from is mostly about processing of tabular/structured data,
our approach generalizes to arbitrary types of data (e.g., images, text, speech).

We filter data assets with less than 50 rows and 2 columns, as they are
unlikely to be used for training models.
\Cref{fig:data-asset-observations} shows two log-transformed density scatter plots
of a de-duplicated subsample of observations.
The first shows the distribution of columns and rows of data assets.
We observe that there are clusters of assets with similar structure.
The second shows the dimension of the data (rows $\times$ columns) and corresponding bytes,
which reveals an (expected) linear relationship, but also a large variability in values.

\iffalse
\begin{figure}[!htpb]%
	\centering
	\subfigure[Columns vs. rows]{%
		\includegraphics[width=0.5\columnwidth]{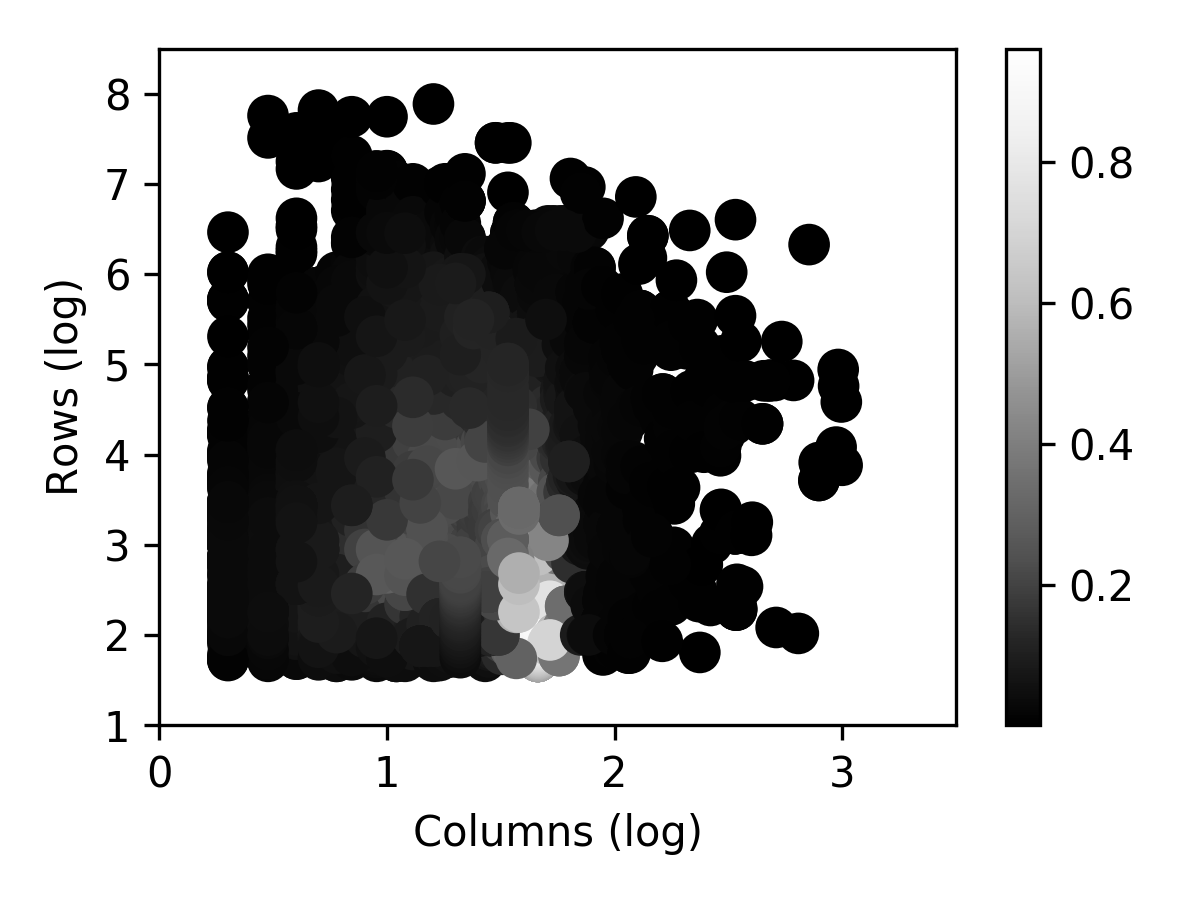}%
	}%
	\subfigure[Data points vs. size in bytes]{%
		\includegraphics[width=0.5\columnwidth]{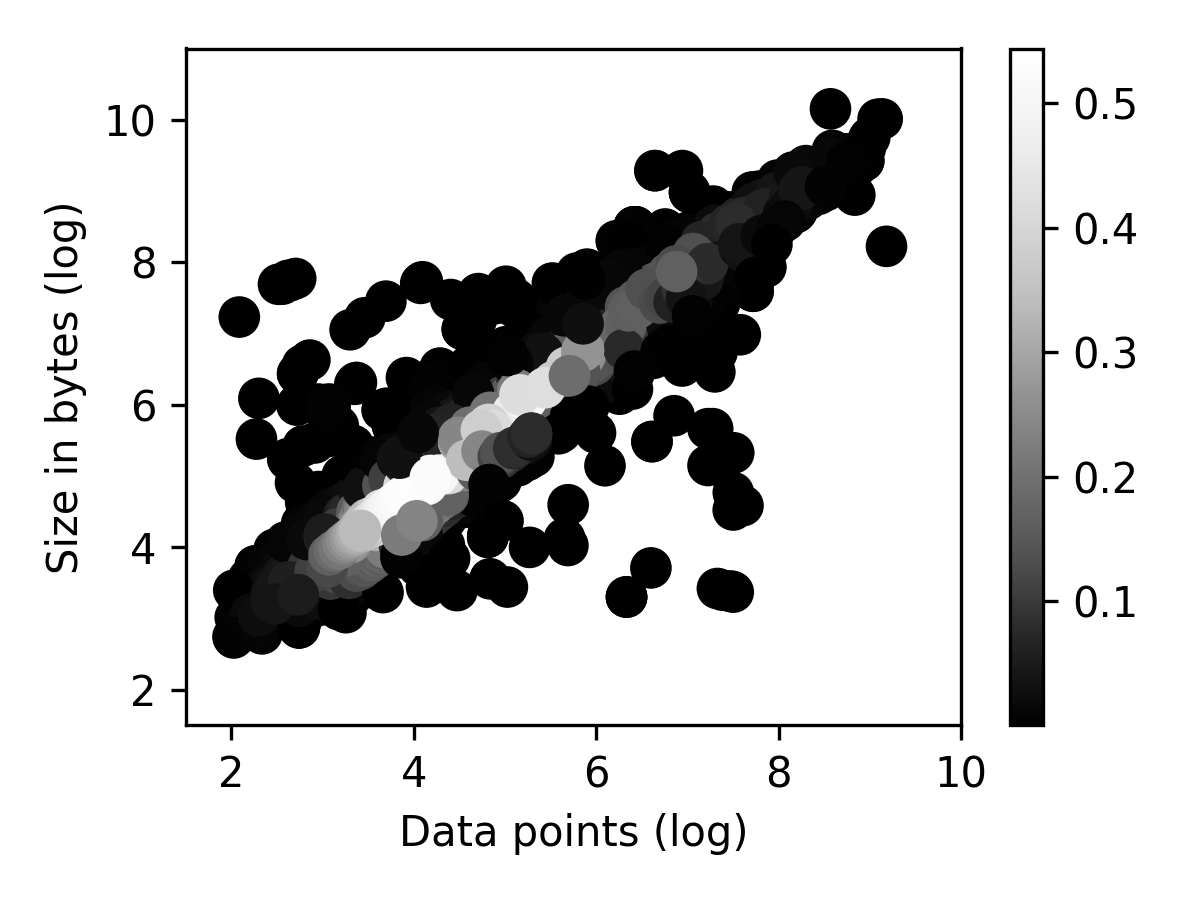}%
	}
	\caption{Observations of data asset dimensions and size in $\mathrm{log}_{10}$-space ($n=9\,821$)}
%	\label{fig:data-asset-observations}
\end{figure}
\else
\begin{figure}[!htpb]%
	\centering
	\includegraphics[width=1\columnwidth]{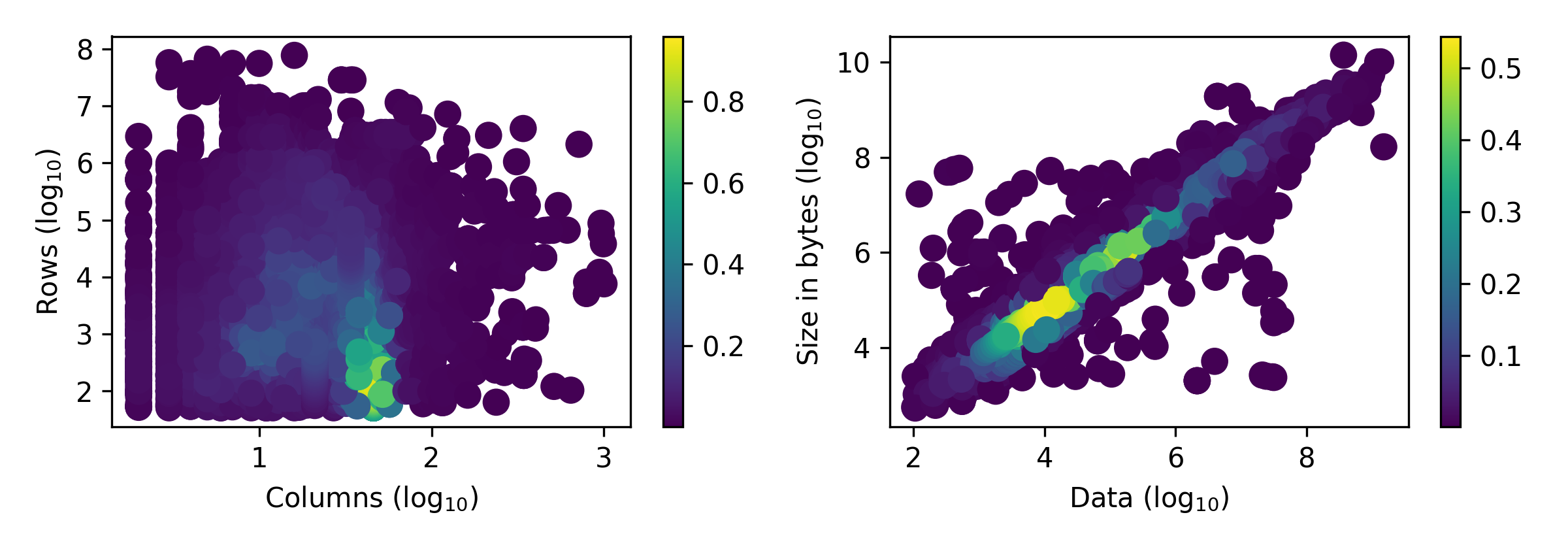}%
	\caption{Observations ($n=9\,821$) of asset dimensions and size}
	\label{fig:data-asset-observations}
\end{figure}
\fi

The model we use to sample a random data asset is a multivariate Gaussian Mixture Model.
We use the scikit-learn implementation to fit a mixture with 50 components and a full covariance matrix on the three-column data set.
The model is then exported into the simulator as described previously.
Because the original data contain many extreme and spread out values which cause too many singletons to form in the mixture, we fit instead on the log-transformed data.
During simulation time, we transform the data back and reject out-of-bound values.

\subsubsection{Simulating Task Execution} \label{sec:impl:simulatetask}

\begin{figure}[!htpb]
	\centering
	\subfigure[Data preprocessing task]{%
		\includegraphics[width=0.5\columnwidth]{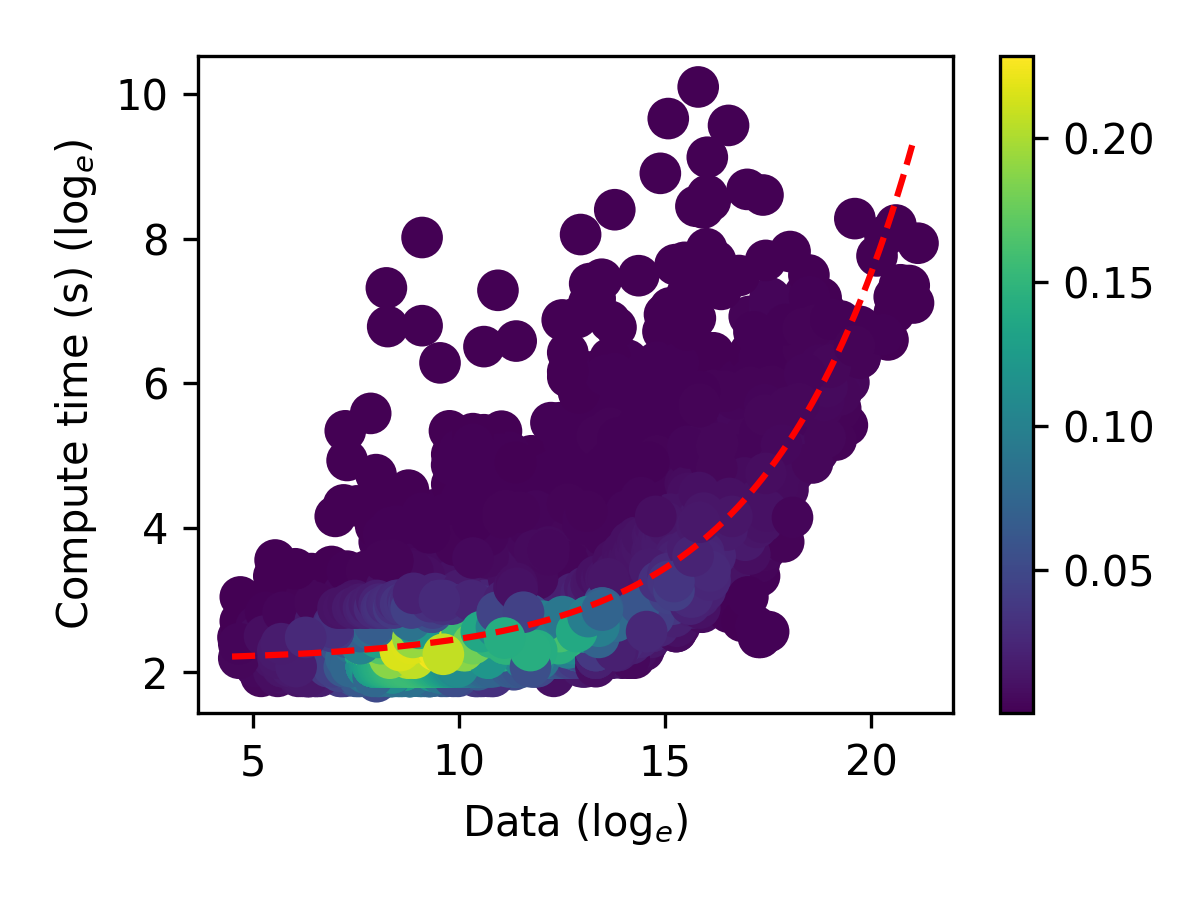}%
		\label{fig:compute-dataprocessing}
	}%
	\subfigure[Training task]{%
		\includegraphics[width=0.5\columnwidth]{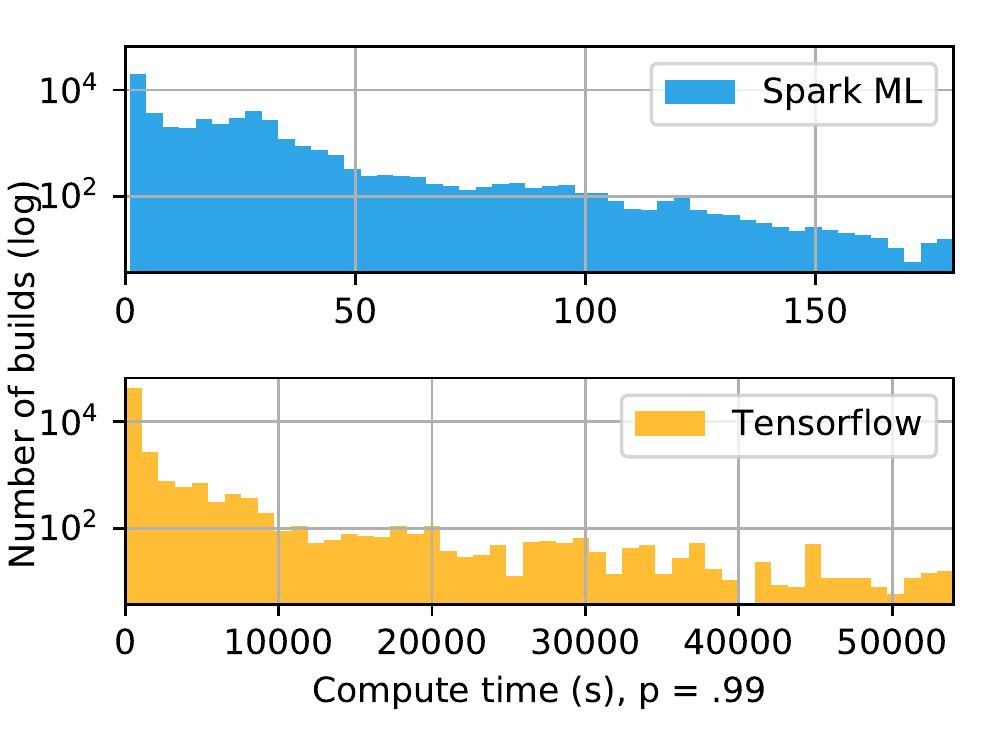}%
		\label{fig:compute-training}
	}
	\caption{Observations of compute time for data preprocessing and training tasks based on other known aspects of the task.}
	\label{fig:compute-time}
\end{figure}

As discussed in \Cref{sec:design:model}, the execution duration of a task typically depends on several factors.
In particular, tasks like data preprocessing and model training are not straight
forward to express analytically, and we instead define statistical models for them.
\Cref{fig:compute-time} shows examples of different ways to simulate the compute
time of data preprocessing and training tasks. Individual plots are explained in more detail below.
%Specifically we describe our model for data processing, training, and validation task compute time estimation.
%All other tasks are currently implemented with placeholder data and modeling them is part of our future work.

\paragraph{Data Preprocessing}

For a data preprocessing task, we can correlate the data asset's size with the computation time.
\Cref{fig:compute-dataprocessing} illustrates this relation.
The red line is an exponential function $f(x) = ab^x + c$
with parameters $a=0.018$, $b=1.330$, $c=2.156$,
fitted on the $\mathrm{log}_e$-transformed data using SciPy's implementation of non-linear least squares.
During simulation time, we use the size of the synthesized data asset to estimate the compute time from the fitted function
and add noise from a log-normal distribution $\alpha = 0.15$, $\mu = -1$
to simulate the long tail around the function.
By associating the data asset $D$ with the task instance $v^p$,
and given the compute resource $R$,
we can express $t(v^p) = t(\mathrm{req}(R)) + t(\mathrm{read}(D)) + t(\mathrm{exec}(v^p, R)) + t(\mathrm{write}(D'))$,
and estimate $t(\mathrm{exec}(v^p, R)) = f(D_d \cdot D_r) + n $ with $n$ being a random sample from our noise function.

Currently we do not model the transformation of $D$ to $D'$, so we simply substitute $D$ for $D'$.
There are other ways to simulate the compute time based on the available data.
For example, sampling from the conditional distribution of compute time given the
data set size from a fitted Gaussian mixture,
or training a different nonlinear regression model and adding noise to the estimate.
We chose the previously described method as we found it produced good results and was straight-forward to implement.
Internals of a preprocessing task, e.g., number and types of operations, will also affect the processing time.
Although we do not have data on this at present, we plan to include it in a future iteration.

\paragraph{Training}

For a training task $v^t$ we know from the assignment we made during pipeline
synthesis which framework $F$ is used, and we know that frameworks have vastly
different execution duration distributions. For example, 50\% of TensorFlow jobs run
in under 180 seconds, whereas 50\% of Spark ML jobs run in under 10 seconds.
\Cref{fig:compute-training} shows two histograms over the compute time of
a subsample ($n=50\,000$) of these jobs.
For a better illustration we only show values below the 99th percentile.
We model $t(v^t)$ in a similar fashion as $t(v^p)$.
To estimate $t(\mathrm{exec}(v^t, R))$ for a given framework $F$,
we stratify the observed execution duration by $F$ and then
fit a Gaussian mixture model $p_F$ on each stratum.
During simulation time we then simply pick a random sample from $p_F$.
This gives us a relatively good fidelity when testing,
e.g., the effect of specific frameworks trending
(which is something we have observed in the production system,
specifically that the number of Tensorflow builds are increasing over time).
In our implementation we also model PyTorch and Caffe training jobs.

\paragraph{Model Evaluation}

For a model evaluation (validation) task $v^e$ we have no data to correlate, only the raw compute time data where we again fit a Gaussian mixture to sample $t(\mathrm{exec}(v^e, R))$ at simulation time.
However, we plan to investigate this further in the next iteration of our simulator.
It is reasonable to assume that the time it takes to validate a model
can be described by the dataset size used for validation,
and the model size (which will generally be a good indicator for the inferencing time).

\paragraph{Model Compression}

Model compression can be used to reduce the size and inferencing time of deep neural networks~\cite{vooturi:17}.
We know from the way state-of-the art model algorithms operate,
that model compression requires roughly as much time as training.
We can therefore re-use the execution duration we have simulated for training
and add Gaussian noise to it to simulate $t(v^c)$. Compression affects several model
metrics, specifically performance, size, and inferencing time.
\Cref{tab:stats-compression} shows these values from experiments we have performed
with GoogleNet and ResNet50 networks on the Food101 training set using Caffe.

\begin{table}[!htpb]
	\caption{Effect of model compression on model parameters}
	\centering
	\begin{tabular}{c | cc | cc | cc }
		\toprule
		Prune &
		\multicolumn{2}{c}{Accuracy (\%)} &
		\multicolumn{2}{c}{Size (MB)} &
		\multicolumn{2}{c}{Inference (ms)} \\
		\midrule
		~ & GN & RN50 & GN & RN50 & GN & RN50\\
		\midrule

		 0 \% & 80.7 & 81.3 & 42.5 & 91.1 & 128 & 223 \\
		20 \% & 80.9 & 80.9 & 28.7 & 83.5 & 117 & 200 \\
		40 \% & 80.0 & 80.8 & 20.9 & 65.2 & 100 & 169 \\
		60 \% & 77.7 & 79.5 & 14.6 & 41.9 &  84 & 141 \\
		80 \% & 69.8 & 69.8 &  8.5 & 8.5  &  71  & 72 \\
		\bottomrule
	\end{tabular}
	\label{tab:stats-compression}
\end{table}

Our simulator currently does not change model metrics for compression tasks in a systematic way,
however we can see that the relative changes in model metrics could be described by a regression model.
Together with a simulation of the run-time view as described in \Cref{sec:design:model},
we could experiment how varying compression levels affect build-time pipeline execution duration compared and run-time inferencing.

\subsubsection{Simulating Pipeline Arrivals}

For simulating pipeline arrivals we model the interarrival in seconds as a random variable and sequentially draw from a fitted distribution.
We collect the timestamps of training job arrivals (which we use as proxy for pipeline arrivals) and calculate the interarrivals.
It is well known that interrarivals typically follow some exponential or related process.
Researchers have found that, for example, TCP traffic is well described by lognormal, Pareto, or Weibull distributions \cite{feldmann:2000}.
On the collected data we found that the exponentiated Weibull distribution produces a good fit.

\begin{figure}[!htpb]
	\centering
	\includegraphics[width=\columnwidth]{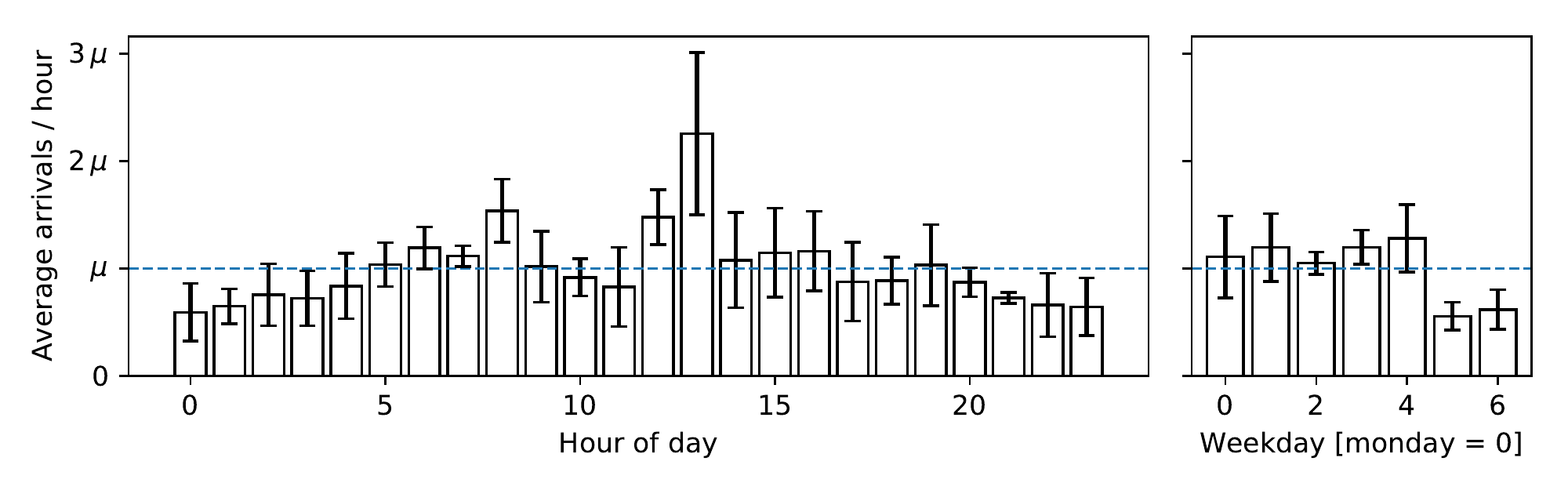}
	\caption{Average arrivals per hour stratified by hour of day and weekday $(n = 210\,824)$. $\mu$ shows the average arrivals per hour aggregated over all values. Error bars show one standard deviation.}
	\label{fig:average-arrivals}
\end{figure}

However, given that humans interact with the system, arrival will differ across different
weekdays and hours of day. \Cref{fig:average-arrivals} shows a histogram of average arrivals per hour,
grouped by hour of day and weekday\footnote{In this paper we report relative numbers for the arrival
rates, as we cannot disclose the absolute numbers by company policy at the present time.
Note that these numbers correspond to the real-world metrics we have collected, and are useful here for illustration purposes.}.
%Specifically it shows that there are fewer training jobs scheduled during the night and weekends.
We can leverage these arrival patterns to predict periods of low infrastructure load for scheduling of automated pipelines.
To provide a realistic arrival profile for our simulation,
we first cluster the calculated interarrivals by hour of day and weekday (resulting in 168 clusters).
On each cluster we fit a log-normal, exponentiated Weibull, and Pareto distribution,
select the best fit based on the sum of square errors (SSE),
and store the parameters together with the hour of day and weekday.
During simulation, we map real timestamps to simulation time, and use that
to sample from the respective cluster.

\subsection{Simulator}

At the core of the experimentation environment lies the simulator.
It implements the conceptual system model and data synthesizers as described in \Cref{sec:design}
as a stochastic, standalone, discrete-event simulator
using the simulation framework SimPy\footnote{\url{https://simpy.readthedocs.io}}.
We also make use of SciPy and scikit-learn for implementing statistical operations such as distribution fitting and sampling.
Synthetic traces are persisted into InfluxDB\footnote{\url{https://www.influxdata.com/time-series-platform/influxdb/}}.
We also developed a Python toolkit for analyzing the experiment data
(which we also use for the evaluation),
and a analytics dashboard using Grafana\footnote{\url{https://grafana.com/}}.
%All code is open source and available on GitHub\footnote{We have open source approval and will include the link to the repository after the reviewing process.}.
We briefly describe how the core concepts are implemented using SimPy.

\paragraph{Resources}

SimPy provides the concept of shared resources\footnote{\url{https://simpy.readthedocs.io/en/latest/topical_guides/resources.html}} to model process interactions.
A shared resource is a congestion point where processes queue up to use them.
We leverage this concept to model our infrastructure.
Each compute resource $R$ (e.g., a training cluster) has an associated job capacity.
When a pipeline task is executed, depending on the task executor implementation, one or more resources for a job will be requested.
If the capacity is reached, the job queues up and waits until a resource is available.
This abstraction is useful because AI ops platforms plug together many different types of infrastructure in a replaceable way,
and therefore generally cannot make assumptions about how system resources behave internally.
For example, details of how a specific training cluster technology provisions workers internally should not leak out to the AI ops platform.
Instead, resources should provide a common interface that allows the platform to reason about a resource's capacity in a high-level way.
SimPy resources provide exactly that.
However, our framework also enables customization of resources, s.t. resource queuing and scheduling mechanisms can be implemented in more detail.

\paragraph{Pipeline Execution}

For simulating pipeline executions as described in \Cref{sec:design:model},
tasks are implemented in plain Python code,
and each system operation $\omega$ is implemented as a SimPy event.
We give an example of how a training task is simulated in this manner.
The training task executor will first attempt to
request the shared resource that models the training cluster $R$.
The queuing time, if any, is used to simulate $\mathrm{req}(R)$.
It then simulates the task execution by generating a timeout event with the value sampled as described in \Cref{sec:impl:stats}.
Afterwards, a Trained Model Asset is created, and properties associated with the model (size, performance, CLEVER score, etc.) are materialized.
For example, to materialize the performance of the model (e.g., the AUC or F-score), we could sample from the distribution of performance values historically observed for the estimator type.
While this is not an accurate estimate of the performance an individual model will have,
it will give us an idea of the overall distribution of, e.g.,
pipelines that may not meet certain quality gates.
The created Trained Model Asset is then persisted to the data store and the execution trace is recorded via the logging interface.

\section{Evaluation}
\label{sec:eval}

To assess the feasibility of our prototype, we evaluate three key aspects.
First, we show how the simulator results can be used to analyze the effect of different parameters on the system.
Second, we examine if the data produced by the simulator reflect the original system under simulation.
Third, we test how well the simulator scales even when simulating years of pipeline executions.

\begin{figure*}[t!]
	\centering
	\includegraphics[width=0.95\textwidth]{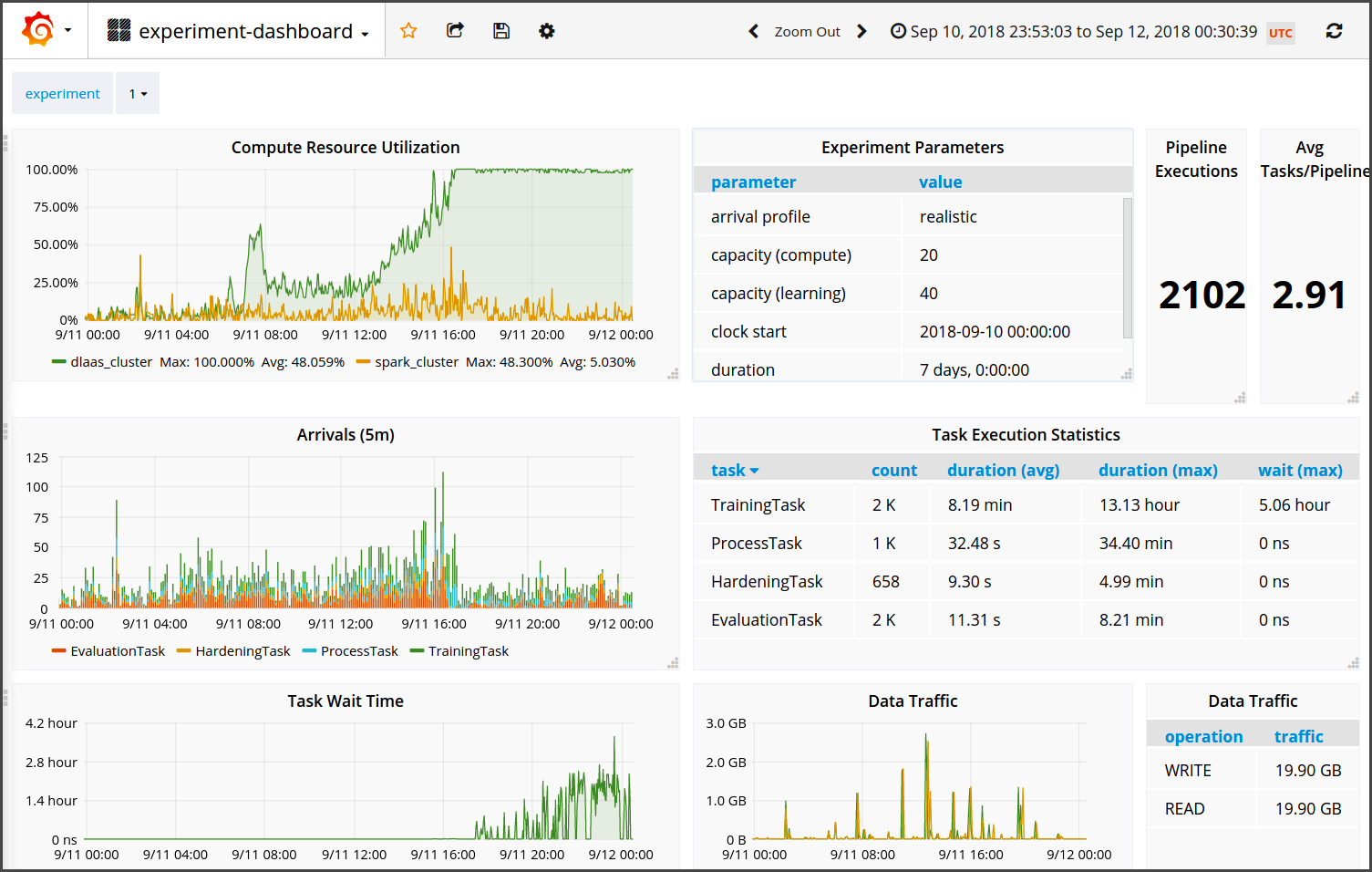}
	\caption{Experiment analytics dashboard showing infrastructure and pipeline execution metrics.}
	\label{fig:dashboard}
\end{figure*}

\subsection{Exploratory Analysis using the Dashboard}

The analytics frontend allows exploratory analysis of experiment results.
\Cref{fig:dashboard} shows the dashboard populated with data from a sample experiment run.
It shows the experiment parameters,
general statistics about individual task executions and wait time.
The graphs shows the resource utilization of compute resources,
individual tasks arrivals, and overall wait time of pipelines, which allows us to
quickly observe the impact of resource utilization on pipeline wait time.
The dashboard also shows the network traffic caused by the execution platform
and includes TCP overhead.

This example illustrates how we can analyze the impact of, e.g., arrival peaks on the infrastructure.
Around 16:00, a typical peak in pipeline arrivals occurs.
The usage of the compute cluster spikes slightly because of preprocessing tasks being scheduled to it.
However, because the learning cluster is quickly saturated by long-running training jobs,
subsequent jobs have to queue,
and post-processing task (like model validation) arrivals are delayed to a later time.
This scenario illustrates how the simulator can be used to examine load balancing of heterogeneous compute infrastructure given different system conditions.

Although the simulation synthesizes and logs model metrics for each pipeline,
we currently do not have a good way of visualizing these data.
We are working on a way to visualize aggregated model metrics (such as overall potential improvement, as described in \Cref{sec:background}) in a meaningful way.
Queries can however be executed, for example, via the InfluxDB web UI, and metrics can be aggregated over pipelines which are identified by unique IDs generated by the system.
This way the lineage of a pipeline can be tracked, and the accuracy over time (which are currently synthesized values and added Gaussian noise) observed.

\subsection{Simulation Accuracy}

\begin{figure}[!htpb]
	\centering
	\scriptsize
	\subfigure[Q--Q plots of task execution duration in $\mathrm{log}_{10}$ seconds]{%
		\includegraphics[width=1\columnwidth]{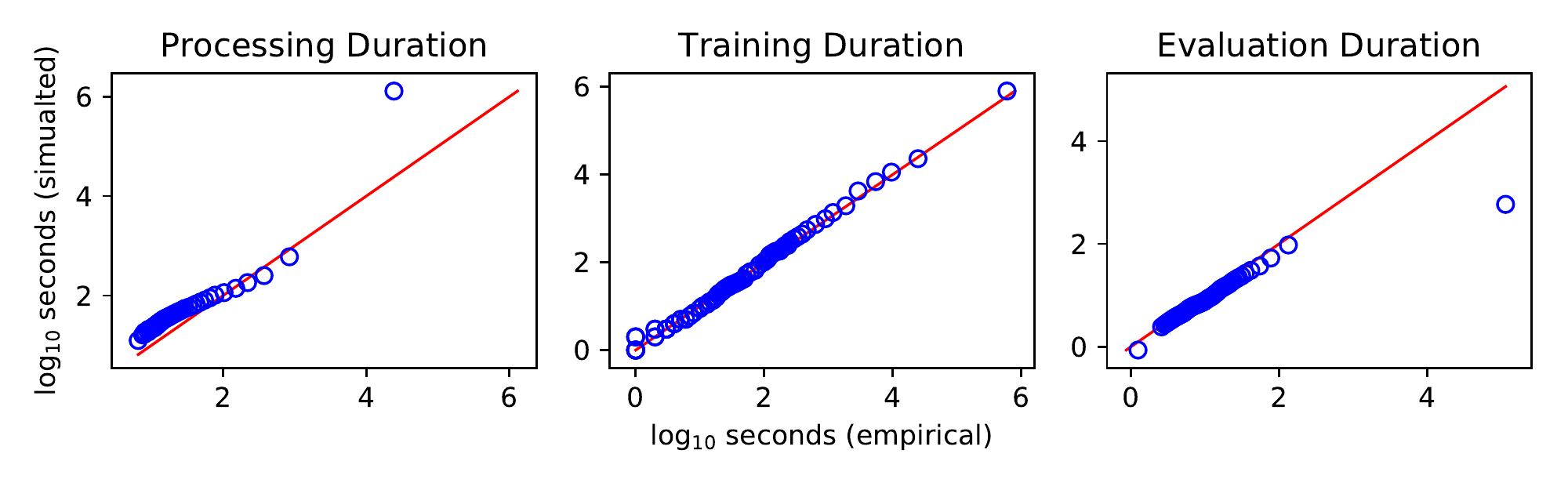}
		\label{fig:eval-duration}
	}
	\subfigure[Q--Q plots of interarrivals in $\mathrm{log}_{10}$ seconds]{%
		\includegraphics[width=1\columnwidth]{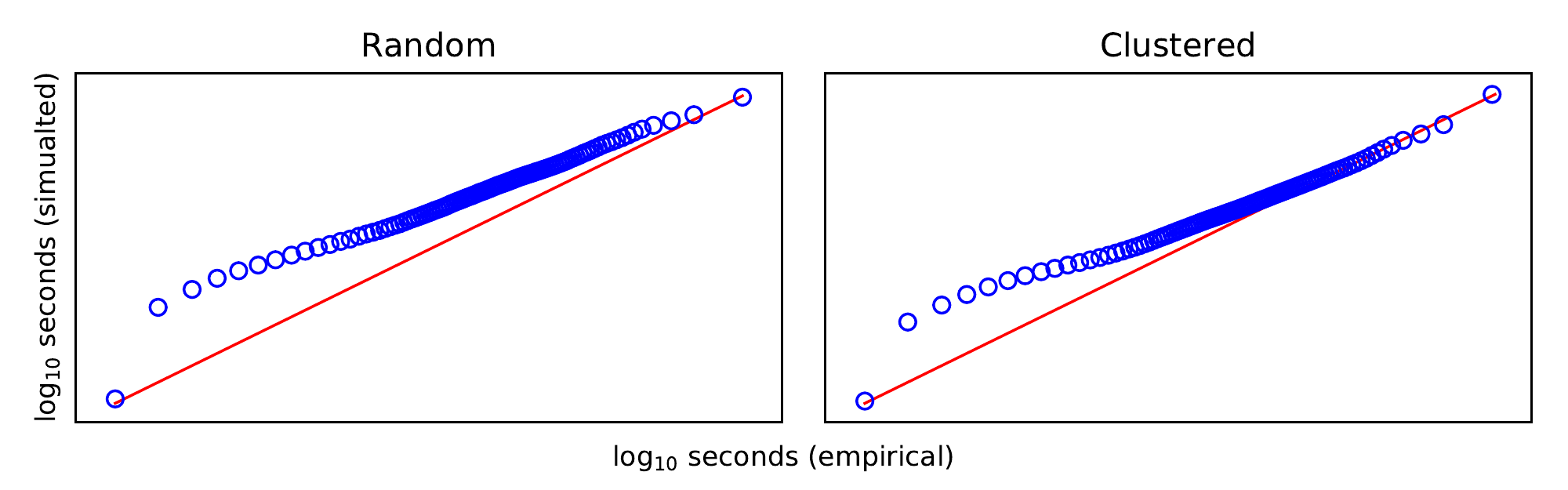}
		\label{fig:eval-arrival-profiles}
	}
	\subfigure[Average arrivals per hour with realistic arrival profile]{%
		\includegraphics[width=1\columnwidth]{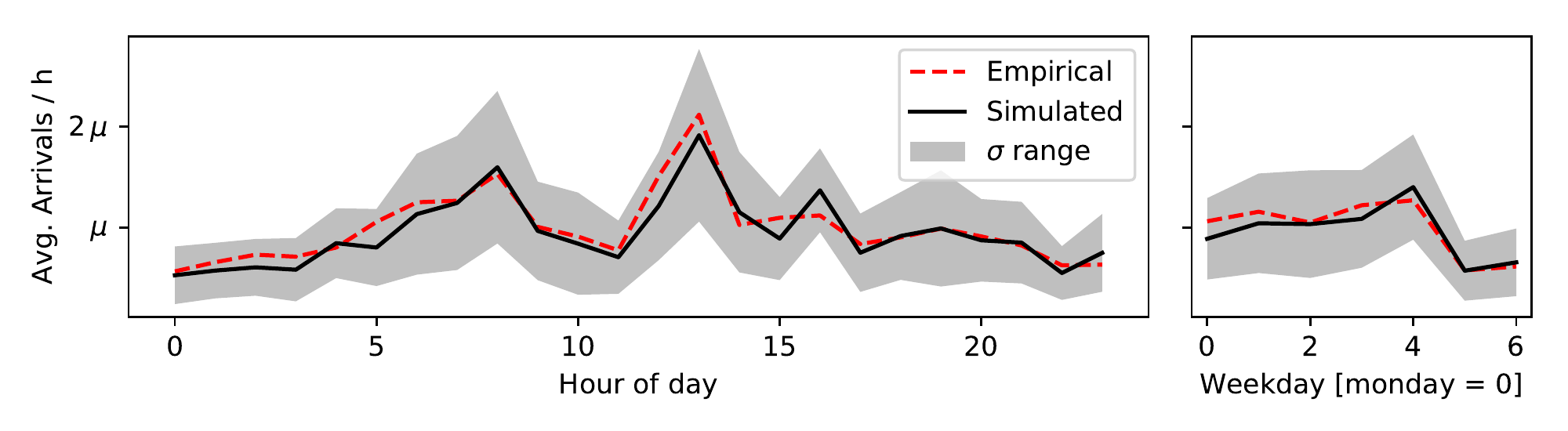}
		\label{fig:eval-arrival-clustered}
	}
	\caption{Statistical evaluation of simulated processes.}
	\label{fig:eval}
\end{figure}

We use the statistical analysis component of \toolname to evaluate the accuracy of our simulation data.
\Cref{fig:eval} shows the results of several pipeline execution experiments to compare
how well the empirical and simulated data agree.

The Q--Q plots in \Cref{fig:eval-duration} plot the quantiles of different task execution durations against each other.
Our preprocessing task simulation slightly overestimates execution duration for short running tasks,
but overall performs well considering the relatively simple statistical model for this complex process.
Training tasks, which we simulated separately for each framework exhibit a very good fit,
which we attribute to the performance of Gaussian mixtures given the large amount of data we have for each framework.
The plot for the evaluation task shows how extreme outliers are sometimes difficult to simulate correctly.

%\iffalse
%\todo{we could remove 11.b if we need space}
The Q--Q plots of interarrivals in \Cref{fig:eval-arrival-profiles} show that both the random
and realistic (clustered by weekday and hour of day) arrival profiles slightly overestimates pipeline interarrivals.
This is acceptable because the experiment environment takes an interarrival factor parameter
that allows us to increase or decrease the average arrivals of pipelines,
and control for such errors.
%\fi
\Cref{fig:eval-arrival-clustered} shows a detailed view of the realistic arrival profile,
where we simulated four weeks of pipeline executions
The black line shows average arrivals per hour in our simulation,
and the red line shows the values from our real system.
We can see that our clustered sampling approach generally performs well in capturing different arrival peaks.

%Additional statistical data on the fitness of our interarrival models can be found in the Appendix.

\subsection{Simulation Performance}

\begin{figure}[!htpb]
	\centering
	\includegraphics[width=\columnwidth]{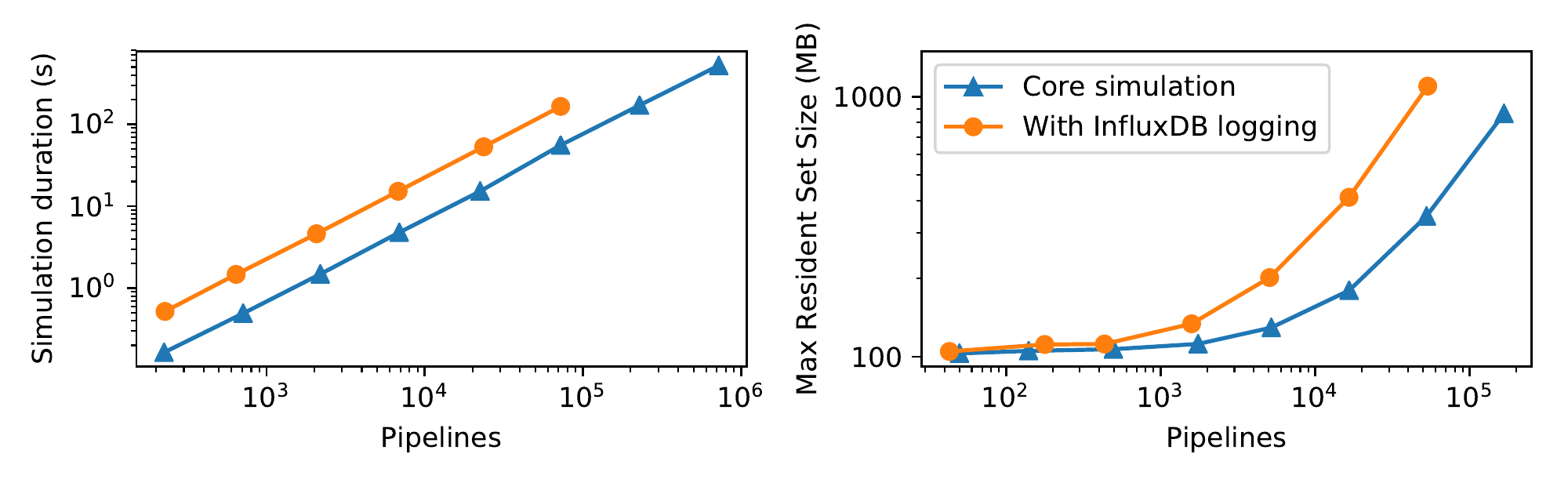}
	\caption{Simulator performance depending on number of pipeline executions}
	\label{fig:simulator-performance}
\end{figure}

To asses the scalability of the simulator,
we run several experiments with an increasing number of pipeline executions and record the (wall clock) execution duration, and memory consumption of the executing python process.
The simulation runs in a single thread and is executed on an AMD FX-8350 4.0GHz CPU machine with 8GB RAM running Linux Mint 18.2 and Python 3.6.
Our experiment results in \Cref{fig:simulator-performance} reveal
a straight-forward linear relationship between the number of executed pipelines and the execution duration,
and likely polynomial memory usage due to the internal storage of execution traces.

We simulated the system for up to a year (365 days), with an average interarrival of 44 seconds,
which corresponds to about 720\,000 pipeline executions.
This simulation took on average 517 seconds, or 8.6 minutes, meaning that simulating
a pipeline execution takes, on average, around 1.4 ms on the evaluation machine.
We ran these experiments several times and observed negligible variance.
The maximum memory consumption for this run was roughly 850 MB.
While we found that the simulator is overall capable of running extensive simulations on a single machine,
we quickly ran into memory issues with InfluxDB when storing more than a few hundred thousand pipelines
due to the way it manages indexes for group-by queries internally.
In fact we were unable to complete the last experiments where we simulated over 100\,000 executions.
While InfluxDB provided an easy-to-use database for rapid prototyping,
we conclude that it was overall a poor choice for the experimentation environment moving forward,
and we will investigate alternatives to maintain better scalability.

\section{Conclusion}
\label{sec:concl}

Efforts of both research and industry to build platforms for democratizing and
operationalizing AI have revealed exciting new opportunities for operations research.
The development of such platforms and their operational strategies is challenging
due to the complex nature of the AI application lifecycle, as well as
the growing need for reconciling build-time and run-time aspects of ML models.

Being able to examine and experiment with the behavior of these systems is a critical
requirement for AI platform operators as our understanding of the AI lifecycle
continues to grow. To that end, we have presented an experimentation and analytics
framework for large-scale AI platforms. Based on our current knowledge of
production-grade AI systems, we developed a system and process model to simulate,
monitor, and analyze the operation of such platforms, enabling continuous
improvement with humans in the loop.

We have shown how to build statistical models from empirical data that accurately
reflect the effect of AI pipeline executions and model scorings on the infrastructure
as well as metrics of the model itself. Simulating a year's worth of pipeline executions
takes only a few minutes on a single machine, enabling a quick feedback cycle to run
experiments. Our evaluation demonstrates that the analytics components can be used to
examine the impact of different system parameters
and support advanced techniques like capacity planning and resource optimization.
For our future work we plan to further extend the current simulator and establish a
stronger link with the run-time of the real system. We envision a mode of operation
where the simulation automatically reconciles its predictions with the real system,
and dynamically adjusts the underlying distributions accordingly, resulting in
an increased fidelity of the simulation. Furthermore, we are working on large-scale
scheduling optimizations for automated retraining pipelines that we
plan to develop and evaluate using our simulator.

\section*{Acknowledgment}

We thank the anonymous reviewers who have contributed to the improvement of this paper.
We also thank Youssef Mroueh, who provided valuable input on the statistical modeling techniques used in our approach.

\bibliographystyle{IEEEtran}
\bibliography{references}

% Generated by IEEEtran.bst, version: 1.14 (2015/08/26)
\begin{thebibliography}{10}
\providecommand{\url}[1]{#1}
\csname url@samestyle\endcsname
\providecommand{\newblock}{\relax}
\providecommand{\bibinfo}[2]{#2}
\providecommand{\BIBentrySTDinterwordspacing}{\spaceskip=0pt\relax}
\providecommand{\BIBentryALTinterwordstretchfactor}{4}
\providecommand{\BIBentryALTinterwordspacing}{\spaceskip=\fontdimen2\font plus
\BIBentryALTinterwordstretchfactor\fontdimen3\font minus
  \fontdimen4\font\relax}
\providecommand{\BIBforeignlanguage}[2]{{%
\expandafter\ifx\csname l@#1\endcsname\relax
\typeout{** WARNING: IEEEtran.bst: No hyphenation pattern has been}%
\typeout{** loaded for the language `#1'. Using the pattern for}%
\typeout{** the default language instead.}%
\else
\language=\csname l@#1\endcsname
\fi
#2}}
\providecommand{\BIBdecl}{\relax}
\BIBdecl

\bibitem{hummer2019modelops}
W.~Hummer, V.~Muthusamy, T.~Rausch, P.~Dube, and K.~El~Maghraoui, ``Modelops:
  Cloud-based lifecycle management for reliable and trusted ai,'' in \emph{2019
  IEEE International Conference on Cloud Engineering (IC2E'19)}, Jun 2019.

\bibitem{wang:17}
K.~Wang, D.~Zhang, Y.~Li, R.~Zhang, and L.~Lin, ``Cost-effective active
  learning for deep image classification,'' \emph{IEEE Transactions on Circuits
  and Systems for Video Technology}, vol.~27, no.~12, pp. 2591--2600, 2017.

\bibitem{zeager:2017}
M.~F. Zeager, A.~Sridhar, N.~Fogal, S.~Adams, D.~E. Brown, and P.~A. Beling,
  ``Adversarial learning in credit card fraud detection,'' in \emph{Systems and
  Information Engineering Design Symposium (SIEDS), 2017}.\hskip 1em plus 0.5em
  minus 0.4em\relax IEEE, 2017, pp. 112--116.

\bibitem{rausch2020pipesim}
T.~Rausch, W.~Hummer, and V.~Muthusmay, ``An experimentation and analytics
  framework for large-scale ai operations platforms,'' in \emph{2020 {USENIX}
  Conference on Operational Machine Learning (OpML '20)}.\hskip 1em plus 0.5em
  minus 0.4em\relax USENIX Association, 2020.

\bibitem{fishman:13}
G.~S. Fishman, \emph{Discrete-event simulation: modeling, programming, and
  analysis}.\hskip 1em plus 0.5em minus 0.4em\relax Springer Science \&
  Business Media, 2013.

\bibitem{simpy:2008}
N.~Matloff, ``Introduction to discrete-event simulation and the simpy
  language,'' \emph{Davis, CA. Dept of Computer Science. University of
  California at Davis. Retrieved on August}, vol.~2, no. 2009, 2008.

\bibitem{keystoneml:2017}
E.~R. Sparks, S.~Venkataraman, T.~Kaftan, M.~J. Franklin, and B.~Recht,
  ``Keystoneml: Optimizing pipelines for large-scale advanced analytics,'' in
  \emph{Data Engineering (ICDE), 2017 IEEE 33rd International Conference
  on}.\hskip 1em plus 0.5em minus 0.4em\relax IEEE, 2017, pp. 535--546.

\bibitem{miao:17}
H.~Miao, A.~Li, L.~S. Davis, and A.~Deshpande, ``Towards unified data and
  lifecycle management for deep learning,'' in \emph{IEEE 33rd International
  Conference on Data Engineering (ICDE)}.\hskip 1em plus 0.5em minus
  0.4em\relax IEEE, 2017, pp. 571--582.

\bibitem{li:2018}
T.~Li, J.~Zhong, J.~Liu, W.~Wu, and C.~Zhang, ``Ease.ml: Towards multi-tenant
  resource sharing for machine learning workloads,'' \emph{Proc. VLDB Endow.},
  vol.~11, no.~5, pp. 607--620, Jan. 2018.

\bibitem{yu:2018}
\BIBentryALTinterwordspacing
C.~Yu, B.~Karlas, J.~Zhong, C.~Zhang, and J.~Liu, ``Multi-device, multi-tenant
  model selection with {GP-EI},'' \emph{CoRR}, vol. abs/1803.06561, 2018.
  [Online]. Available: \url{http://arxiv.org/abs/1803.06561}
\BIBentrySTDinterwordspacing

\bibitem{gama:14}
J.~Gama, I.~{\v{Z}}liobait{\.e}, A.~Bifet, M.~Pechenizkiy, and A.~Bouchachia,
  ``A survey on concept drift adaptation,'' \emph{ACM computing surveys
  (CSUR)}, vol.~46, no.~4, p.~44, 2014.

\bibitem{hind:18}
M.~Hind, S.~Mehta, A.~Mojsilovic, R.~Nair, K.~N. Ramamurthy, A.~Olteanu, and
  K.~R. Varshney, ``Increasing trust in ai services through supplier's
  declarations of conformity,'' \emph{arXiv:1808.07261}, 2018.

\bibitem{tfx:2017}
D.~Baylor, E.~Breck, H.-T. Cheng, N.~Fiedel, C.~Y. Foo, Z.~Haque, S.~Haykal,
  M.~Ispir, V.~Jain, L.~Koc \emph{et~al.}, ``Tfx: A tensorflow-based
  production-scale machine learning platform,'' in \emph{Proceedings of the
  23rd ACM SIGKDD International Conference on Knowledge Discovery and Data
  Mining}.\hskip 1em plus 0.5em minus 0.4em\relax ACM, 2017, pp. 1387--1395.

\bibitem{liu:2012}
Y.~Liu, H.~Zhang, C.~Li, and R.~J. Jiao, ``{Workflow simulation for operational
  decision support using event graph through process mining},'' \emph{Decision
  Support Systems}, vol.~52, no.~3, pp. 685--697, feb 2012.

\bibitem{cloudsim:2011}
R.~N. Calheiros, R.~Ranjan, A.~Beloglazov, C.~A. De~Rose, and R.~Buyya,
  ``Cloudsim: a toolkit for modeling and simulation of cloud computing
  environments and evaluation of resource provisioning algorithms,''
  \emph{Software: Practice and experience}, vol.~41, no.~1, pp. 23--50, 2011.

\bibitem{icancloud:2012}
A.~N{\'u}{\~n}ez, J.~L. V{\'a}zquez-Poletti, A.~C. Caminero, G.~G.
  Casta{\~n}{\'e}, J.~Carretero, and I.~M. Llorente, ``icancloud: A flexible
  and scalable cloud infrastructure simulator,'' \emph{Journal of Grid
  Computing}, vol.~10, no.~1, pp. 185--209, 2012.

\bibitem{Lin:2018}
\BIBentryALTinterwordspacing
J.~Lin, J.~Zhang, Y.~Ding, L.~Zhang, and Y.~Han, ``All-spark: Using simulation
  tests directly in production environments to detect system bottlenecks in
  large-scale systems,'' in \emph{Proceedings of the 19th International
  Middleware Conference}, ser. Middleware '18.\hskip 1em plus 0.5em minus
  0.4em\relax New York, NY, USA: ACM, 2018, pp. 1--12. [Online]. Available:
  \url{http://doi.acm.org/10.1145/3274808.3274809}
\BIBentrySTDinterwordspacing

\bibitem{hernandez:2012}
J.~Hern{\'a}ndez-Orallo, P.~Flach, and C.~Ferri, ``A unified view of
  performance metrics: translating threshold choice into expected
  classification loss,'' \emph{Journal of Machine Learning Research}, vol.~13,
  no. Oct, pp. 2813--2869, 2012.

\bibitem{weng:18}
T.-W. Weng, H.~Zhang, P.-Y. Chen, J.~Yi, D.~Su, Y.~Gao, C.-J. Hsieh, and
  L.~Daniel, ``Evaluating the robustness of neural networks: An extreme value
  theory approach,'' \emph{arXiv preprint arXiv:1801.10578}, 2018.

\bibitem{tramer:16}
F.~Tram{\`e}r, F.~Zhang, A.~Juels, M.~K. Reiter, and T.~Ristenpart, ``Stealing
  machine learning models via prediction apis.'' in \emph{USENIX Security
  Symposium}, 2016, pp. 601--618.

\bibitem{bifet:15}
A.~Bifet, G.~de~Francisci~Morales, J.~Read, G.~Holmes, and B.~Pfahringer,
  ``Efficient online evaluation of big data stream classifiers,'' in
  \emph{Proceedings of the 21th ACM SIGKDD international conference on
  knowledge discovery and data mining}.\hskip 1em plus 0.5em minus 0.4em\relax
  ACM, 2015, pp. 59--68.

\bibitem{sethi:18}
T.~S. Sethi and M.~Kantardzic, ``Handling adversarial concept drift in
  streaming data,'' \emph{Expert Systems with Applications}, vol.~97, pp.
  18--40, 2018.

\bibitem{sun:16}
Y.~Sun, Z.~Wang, H.~Liu, C.~Du, and J.~Yuan, ``Online ensemble using adaptive
  windowing for data streams with concept drift,'' \emph{International Journal
  of Distributed Sensor Networks}, vol.~12, no.~5, p. 4218973, 2016.

\bibitem{demvsar:18}
J.~Dem{\v{s}}ar and Z.~Bosni{\'c}, ``Detecting concept drift in data streams
  using model explanation,'' \emph{Expert Systems with Applications}, vol.~92,
  pp. 546--559, 2018.

\bibitem{feldmann:2000}
A.~Feldmann, ``Characteristics of tcp connection arrivals,'' \emph{Self-Similar
  Network Traffic and Performance Evaluation}, pp. 367--399, 2000.

\bibitem{vooturi:17}
D.~T. Vooturi, S.~Goyal, A.~R. Choudhury, Y.~Sabharwal, and A.~Verma,
  ``Efficient inferencing of compressed deep neural networks,'' \emph{CoRR},
  vol. abs/1711.00244, 2017.

\end{thebibliography}

\end{document}